\documentclass[twocolumn,preprintnumbers,superscriptaddress,amsmath,amssymb,prb]{revtex4}

\usepackage{graphicx}
\usepackage{dcolumn}
\usepackage{subfigure}
\usepackage{bm}
\usepackage[utf8]{inputenc}  
\usepackage{comment}

\usepackage[cmyk,dvipsnames]{xcolor}
\definecolor{pacificb}{HTML}{1CA9C9}

\def \average#1{\left\langle #1 \right\rangle}
\def \roundb#1{\left( #1 \right)}

\def \squareb#1{\left[ #1 \right]}
\def \be {\begin{equation}}
\def \ee {\end{equation}}
\def \beA {\begin{eqnarray}}
\def \eeA {\end{eqnarray}}
\def \ua {\uparrow}
\def \da {\downarrow}

\def \Tr  {\mbox{Tr}}

\def \Im  {\mbox{Im}}
\def \be{\begin{equation}}
\def \ee{\end{equation}}

\def\u{\uparrow}
\def\d{\downarrow}
\def\Tr{\mbox{Tr}}

\begin{document}

\title{Micromagnetic simulations of spin-torque driven magnetisation dynamics with spatially resolved spin transport and magnetisation texture}

\author{Simone Borlenghi} 
\affiliation{Department of Physics and Astronomy, Uppsala University, Box 516, SE-75120 Uppsala, Sweden.}
\author{M. R. Mahani}
\affiliation{Department of Applied Physics, School of Engineering Sciences, KTH Royal Institute of Technology, Electrum 229, SE-16440 Kista, Sweden}
\author{Hans Fangohr}
\affiliation{University of Southampton, Southampton, SO17 1BJ, United Kingdom}
\affiliation{European XFEL GmbH, Holzkoppel 4, 22869 Schenefeld, Germany}
\author{M. Franchin}
\affiliation{Cambridge, United Kingdom}
\author{Anna Delin}
\affiliation{Department of Applied Physics, School of Engineering Sciences, KTH Royal Institute of Technology, Electrum 229, SE-16440 Kista, Sweden}
\affiliation{Department of Physics and Astronomy, Uppsala University, Box 516, SE-75120 Uppsala, Sweden.}
\affiliation{Swedish e-Science Research Center (SeRC), KTH Royal Institute of Technology, SE-10044 Stockholm, Sweden}
\author{Jonas Fransson}
\affiliation{Department of Physics and Astronomy, Uppsala University, Box 516, SE-75120 Uppsala, Sweden.}

\begin{abstract}
We present a simple and fast method to simulate spin-torque driven magnetisation dynamics in nano-pillar spin-valve structures. The approach is based on
the coupling between a spin transport code based on random matrix theory and a micromagnetics finite-elements software. In this way the spatial dependence
of both spin transport and magnetisation dynamics is properly taken into account. Our results are compared with experiments. The excitation
of the spin-wave modes, including the threshold current for steady state magnetisation precession and the nonlinear frequency shift of the modes are
reproduced correctly. The giant magneto resistance effect and the magnetisation switching also agree with experiment. The similarities with recently
described spin-caloritronics devices are also discussed.
\end{abstract}

\date{\today}
\maketitle

\section{Introduction}%

The orientation of the magnetization in a magnetic film can be influenced using a spin-polarized current. Consequently, a direct current can transfer spin angular momentum between magnetic layers, separated by either a normal metal or a thysichin insulating layer. This effect is called spin transfer torque (STT) and was first discussed in the 1970s in the context of moving magnetic domain walls \cite{berger78} and fully understood in the 1990s \cite{berger96,slonczewski96}. STT has been of profound importance for the development of spintronic devices such as read-heads based on the giant magnetoresitive (GMR) effect \cite{baibich88,binash89}, the spin-transfer torque magnetic random-access memory (STT-MRAM) \cite{diao07} and spin-torque nano-oscillators (STNO)\cite{katine00}.

Until very recently \cite{abert14} the approaches to theoretically describe the magnetization dynamics induced by a spin torque usually greatly simplified or neglected the description of either the spatial inhomogeneity of the spin torque, or the three-dimensional magnetization texture \cite{lee04,xiao05,berkov08}.

In the present work, we go beyond such approaches by coupling a finite element micromagnetic method \cite{fischbacher07} to a numerical solver for spin transport, based on continuous random matrix theory (CRMT) \cite{rychkov09,borlenghi11}.
In this way the effect of spin torque is described, with the transport and magnetic degrees of freedom treated on an equal footing. The spatial inhomogeneity of both spin transport and magnetization dynamics is thus explicitly included. In our implementation, CRMT is parametrised by the same set of experimentally accessible parameters as in Valet-Fert theory \cite{valet93}, so that our numerical simulations contain no free parameters.

We demonstrate the capabilities of our computational method by addressing theoretically the effect of spin torque on the magnetization dynamics in the perpendicularly magnetized circular spin-valve nanopillar experimentally investigated in Ref.~[\onlinecite{naletov11}]. This configuration is obtained by saturating the device with a large applied field perpendicular to the layers. The device setup is kept very simple and as such serves as a prototype for spin-valve structures, which find application also in the emerging field of spin-caloritronics \cite{borlenghi14a,borlenghi14b,borlenghi15a,borlenghi15b}. In particular, our configuration corresponds to a circular precession of the magnetisation, allowing for a precise identification of the spin wave (SW) modes.This is crucial if one wants to couple the system to an external rf signal, since only signal with the same symmetries of the SW modes can excite the magnetisation. Breaking the axial symmetry 
results in a more complicated configuration where modes with different symmetries mix up \cite{naletov11}.

Moreover, the system can be described using the language of coupled oscillators \cite{slavin09} and in particular of the discrete nonlinear Schr\"odinger equation \cite{borlenghi14a} (DNLS). Indeed, this setup corresponds to the simplest realisation of the DNLS, containing only two elements. Here spin transfer torque physically corresponds to a magnon chemical potential \cite{borlenghi14b} that controls the propagation of the energy and magnetisation currents between the two layers.
The DNLS appears in many branches of Physics, including Bose-Einstein condensates, photonics waveguides and photosynthetic reactions. Understanding the dynamics in our setup can therefore shed light 
on a very general oscillator model.


The remainder of this paper is organized as follows: 
in Sec.~II we describe the geometry of the nanopillar and also briefly review the classification of SW modes in a perpendicularly magnetised nanopillar. Those modes are then identified by means of micromagnetic simulations at zero current
(thus without spin torque). This study was previously performed in Ref.~[\onlinecite{naletov11}], to which we refer for a thorough discussion.
Sec. III reviews key concepts of scattering approach and CRMT to describe transport in magnetic multilayers, and consitutes an extension of the material presented in Refs. [\onlinecite{waintal00}] and [\onlinecite{rychkov09}].
In Sec.~IV we describe how to couple the CRMT transport code to Nmag \cite{fischbacher07}, in order to simulate the effect of spatially inhomogeneous spin transfer torque and magnetisation dynamics on the same footing. 
Sec.~V contains our micromagnetic simulations of current driven spin dynamics. Here we identify the SW modes excited by spin transfer torque, the critical current for auto oscillations, and the frequency shift beyond the critical threshold \cite{slavin09}.
Using CRMT we provide a precise characterisation of the magnetoresistance. Our simulations are then compared with the experimental results found in Ref.~\onlinecite{naletov11}.
Finally, in the conclusion we summarise the main results of this paper and point out further possible developments. 

\section{Physical system and model}%

\subsection{Spin valve structure}%

\begin{figure}
\begin{center}
\includegraphics[width=9.0cm]{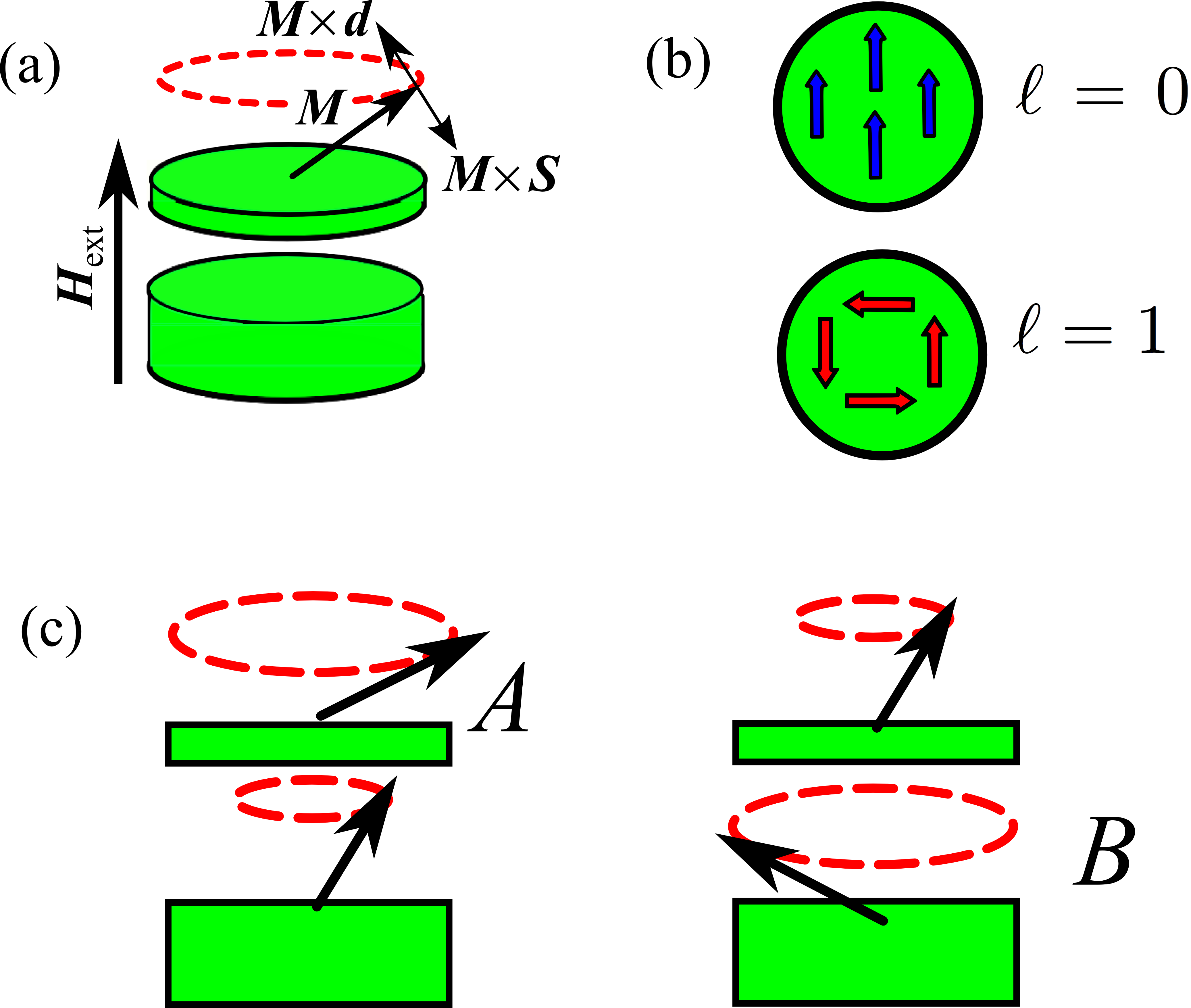}
\caption{(Color online) a) Sketch of the nano-pillar under study, with the arrows representing the magnetisation vectors.   In the steady state, the spin transfer torque $\bm{M}\times\bm{S}$ compensates the damping torque $\bm{M}\times\bm{d}$ and the local magnetization vector $\bm{M}_s$ precesses regularly at the Larmor frequency along a circular orbit. (b) The SW modes in each disk are Bessel functions $J_{\ell m}$, with ($\ell$) ($m$) the azimuthal (radial) index. Modes with $\ell=0$ correspond to a uniform in-phase precession of the magnetisation, while modes with $\ell=1$ correspond to a non-uniform precession, with the magnetisation vector rotating along the azimuthal direction.
c) In the case where two disks are coupled through the dipolar interaction, the SW $J_{\ell m}^{B/A}$ modes separate into binding (B) and anti-binding (A), which correspond, respectively, to an anti-phase precession 
(occurring mainly in the thick layer) and in-phase precession (occurring mainly in the thin layer).}
\label{fig:figure1}
\end{center}
\end{figure}

The nanopillar studied here is displayed in Fig.\ref{fig:figure1}(a). It consists of a trilayer structure made of two Permalloy ($\rm{Ni_{80}Fe_{20}}$  alloy) disks Py$_a$ and Py$_b$ separated by a 10 nm Cu spacer. The disks have diameter $d=200$~nm and thicknesses $t_a=4$~nm (upper disk) and $t_b=15$~nm (lower disk). The upper disk is connected to a 25~nm Au contact and the lower disk to a 60~nm Cu contact.
In the experiment\cite{naletov11}, the sample was mounted inside a magnetic resonance force microscope (MRFM) and the whole apparatus was placed inside a vacuum chamber operated at room temperature. The external magnetic field $\bm{H}_{\rm{ext}}$, was oriented along the pillar axis $\bm{z}$, which corresponds to the precession axis of the magnetisation.

The MRFM consists of an ultra-soft cantilever with a 100 nm diameter magnetic sphere glued to its tip. The sphere is positioned precisely above the center of the nano-pillar, so as to retain the axial symmetry.
The mechanical-FMR spectroscopy consists in recording, by optical means, the vibration amplitude of the cantilever as a function of the bias out-of-plane magnetic field in the presence of a RF field with fixed frequency excitation \cite{naletov11}. The dynamics can be excited also by injecting a dc current along $\bm{z}$, which excites the dynamics in the thin layer due to spin transfer torque.
In the steady state, the combined torques exerted by the RF excitations and spin transfer compensate the damping, so that the local magnetization vector precesses regularly at the Larmor frequency along a circular orbit, see Fig.\ref{fig:figure1}a). 

The instantaneous magnetization $\bm{M}$ can be decomposed into a large static component $\bm{M}_\parallel$ and a small oscillating component $\bm{M}_\perp$, where $\bm{M}_\parallel$ is parallel to the local precession axis (i.e., the direction of the external magnetic field), and $\bm{M}_\perp$ is perpendicular to that axis. 

An essential feature of our system is that different SW modes have different spatial distribution of the phase of $\bm{M}_\perp$ inside the magnetic disks. Those modes can be excited only by RF fields with the same rotational symmetries, giving \emph{selection rules} for the excitation of the SW modes \cite{naletov11}.

The dipolar force acting on the cantilever is proportional to the spatially averaged value of the longitudinal (static) component of the magnetization
inside the whole nanopillar, 
\be\label{eq:mzav}
\average{M_z}=\frac{1}{V}\int_VM_z(\bm{r})d^3\bm{r}.
\ee 
The latter is not subject to any selection rule, so that the the mechanical-FMR setup detects all possible SW modes. Experimentally, it has been observed that the presence of the cantilever introduces
a shift of $+0.57$ GHz in the SW spectrum. This has be taken into account in our simulations.

The dynamics of the magnetization can be excited also by means of an rf current flowing along the axis of the pillar, which generates an rf orthoradial
Oersted field, and by mean of an homogeneous in-plane RF magnetic field.

We remark that, although in experiments the dynamics driven by a combination of RF fields and STT, in our simulations the dynamics is excited by STT only. 
The effect of the RF field is modelled by setting different initial condition for the magnetisation. 

In our circuit, a positive current corresponds to a flow of electrons from the bottom Py$_b$ thick layer to the top Py$_a$ thin layer, 
and stabilizes the parallel configuration due to the spin transfer effect. 
Vice-versa, a negative current stabilizes the thick layer and destabilizes the thin one. At low current, thick and thin layer are thus the fix and free layer correspondingly.
However, at high enough current, both layers precess due to the repulsive dipolar interaction, as will be discussed later.

\subsection{Magnetization dynamics} %
In this section, we briefly review the magnetization dynamics  in our system. For a more comprehensive discussion, see Ref. [\onlinecite{naletov11}].

The local dynamics of the magnetisation $\bm{M}^j$, which depends continuously on the position $\bm{r}^j$ in the layer $j=(a,b)$,
is described by the Landau-Lifshitz-Gilbert equation \cite{gilbert55,landau65}:

\be\label{eq:llg1}
\frac{1}{\gamma}\dot{\bm{M}^j} = \bm{M}^j \times \bm{H}_{\text{eff}}^j
+ \alpha\bm{M}^j \times \dot{\bm{M}}^j +  \bm{M}^j \times \bm{S}^j.
\ee
Here, $\gamma<0$ is the gyromagnetic ratio in the magnetic layer.  The first term on the right-hand side of Eq.(\ref{eq:llg1}) describes the
adiabatic torque, that accounts for the precession of the magnetization vector around the local equilibrium direction. This precession axis is defined
by the effective magnetic field experienced locally by the magnetization, $\bm{H}^j_{\text{eff}}= -\frac{\partial F}{\partial{\bm{M}}^j}$, which
contains all the static contributions to the free energy $F$ of the layers \cite{gurevich96}. In particular, the effective field contains contributions from applied field, exchange interaction, and dipolar interaction
between the layers. We refer to Refs.\onlinecite{gurevich96,naletov11,borlenghi15b,borlenghi11} for the explicit expressions.

The second terms on the right-hand side of Eq.(\ref{eq:llg1}) is the damping torque
 
\be\label{eq:damp}
\bm{d}^j=\alpha^j\dot{\bm{M}}^j
\ee
proportional to the Gilbert damping parameter $\alpha^j$.
We introduce here the notation $\bm{M}^j \equiv M_s^j\bm{m}^j$, with $M_s^j$ the norm of the magnetization (a constant of the motion) 
and $\bm{m}^j$ the unit vector along the magnetization direction.  The dissipative term $\bm{d}^j$ is responsible for the finite linewidth (full width at half height) 
of the resonance peaks, $\Delta H=2 d$. For a normally magnetized nanopillar, with circular precession of the magnetization,  
the simple relation $\alpha=|\gamma| \Delta H/ (2\omega)$ holds \cite{slavin09}.

If a charge current $I_{dc}$ is flowing between two layers $j$ and $j^\prime$, the Slonczewski-Berger \cite{slonczewski96,berger96} spin
transfer torque reads

\be\label{eq:stt}
\bm{S}^j=\frac{I_{dc}}{2\pi\lambda}\squareb{{\bm{m}^j} \times{\bm{m}^{j\prime}}}.
\ee
The latter depends on the relative angle between the magnetization $\bm{m}^{j}$ in the layer $j$ and the spin
polarization of the current, which coincides with the direction of the magnetization $\bm{m}^{j^\prime}$ of the polarizer (here the thick layer). The term 
\be
\lambda^j=\frac{2 e M_s^j V}{\eta h}
\ee
has the dimension of a distance. Here, $h$ is the Planck constant, $e$ the absolute value of the electron charge. 
$\eta$ is the spin polarisation of the current and $V$ the volume of the thin layer. 

Since $\bm{d}^j$
and $\bm{S}^j$ are collinear, spin transfer torque can compensate the damping torque, as shown in Fig.\ref{fig:figure1}. 
When the dc current through the nano-pillar reaches the threshold current $I_{\rm{th}}=-2 \pi \lambda \alpha H_{\rm{eff}}$, the
thin layer starts auto-oscillating.
Combining Eqs.(\ref{eq:damp}) and (\ref{eq:stt}), it is possible to define an effective damping for the thin layer, 
\be\label{eq:effdamp}
d=\alpha(1-I_\text{dc}/I_\text{th}),
\ee
which depends linearly on the spin polarized dc current \cite{sankey06,chen08}. The critical thresohold corresponds to the value of
the current $I_{dc} $ at which the effective damping vanishes and the system starts auto-oscillating.

\subsection{Coupled oscillator model and classification of the SW modes}%
Since our layers are thinner than 15 nm, one can assume that the magnetization dynamics is uniform along the thickness. In this
approximation, the linearized LLG equation simplifies to two equations describing the
circular precession of the transverse magnetization projections $M_x^j$ and $M_y^j$ around the $z$ axis, which depends only on the two spatial variables $(x,y)$
in the layer $j$ \cite{gurevich96,naletov11}. The two real equations of each layer can be rewritten as one complex equation for the dimensionless spin-wave amplitude 

\be\label{eq:swamplitude}
c_j=\frac{M^j_x + i M^j_y}{\sqrt{2M^j_s(M^j_s+M^j_z})},
\ee
that depends on the polar coordinates $(r_j,\phi_j)$ of disk $j$. 

The dynamics of the two disks, written in terms of the $c_j$s, is described by the equations \cite{slavin09,naletov11,borlenghi14a}
\beA
\dot{c}_a &=& i\omega_ac_a-[\Gamma_{a-}-\Gamma_{a+}]c_a+ih_{ab}c_b\label{eq:ca}\\
\dot{c}_b &=& i\omega_bc_b-[\Gamma_{b-}-\Gamma_{b+}]c_b+ih_{ba}c_a\label{eq:cb},
\eeA
which are the equations of motion of two coupled nonlinear oscillators with resonance frequency $\omega_j (p_j)$ and damping rates $\Gamma_j(p_j)$. 
Both depend on the SW power $p_j=|c_j|^2$, which describes the amplitude of the oscillations in each disk.  From hereon, to keep the notation simple we do not write explicitly the dependence on $p_j$.
 The frequencies $\omega_j=\gamma |H^j_{\rm{eff}}|$
are proportional to the local magnetic field, while the damping rates $\Gamma_{j-}$ are proportional to $\alpha_j\omega_j$.
Both can be therefore controlled by means of the applied field along $\bm{z}$. 

The terms $\Gamma_{j+}$, proportional to $I_{dc}$ are due to spin transfer torque, which can compensate the damping and lead to auto oscillations of the layers.
In the present case, those terms do not have the same sign. At positive current, $\Gamma_{a+}$ is positive, while $\Gamma_{b+}$ is negative,
favouring the auto oscillations in the thin layer $a$, and stabilising the thick layer $b$. 

The dipolar coupling strength $h_{jj^\prime}$ is an effective term obtained by averaging the dipolar field over the volumes of the samples, see Refs. \onlinecite{naletov11} for the explicit expression.

Eqs. (\ref{eq:ca}) and (\ref{eq:cb}) describe the dynamics of a nonlinear Schr\"odinger dimer, the simplest realisation of the discrete nonlinear Schr\"odinger equation (DNLS) \cite{iubini13,borlenghi14a,borlenghi14b}. Upon multiplying Eqs.(\ref{eq:ca}) and (\ref{eq:cb}) respectively by $c_a^*$ 
and $c_b^*$ and summing them with their complex conjugate equations, one has the following continuity equation for the SW power

\be\label{eq:continuity}
\dot{p}_a=-2(\Gamma_{a-}-\Gamma_{a+}) p_a+j^p_{ab}.
\ee
and a similar equation for $p_b$. The magnetisation current $j^p_{ab}=2\Im[h_{ab}c_ac^*_b]$ describe the transfer of $M_z$ between the two layers, and is essentially the SW current written for a discrete systems with only two spins \cite{borlenghi14a}.
Upon writing $c_a=\sqrt{p_a(t)}e^{i\phi_a(t)}$, the current reads 

\be\label{eq:swcurrent}
j_{ab}^p=2h_{ab}\sqrt{p_ap_b}\sin[\phi_a(t)-\phi_b(t)+\beta]. 
\ee
The quantity $\beta$ comes from the condition of dissipative coupling between the oscillators \cite{iubini13,borlenghi15a}.
When the two oscillators are synchronised, $\phi_a\approx\phi_b$ and the magnetisation current approaches the constant value $j^p_{ab}\propto\sin\beta$. On the other hand, if the oscillators are not synchronised, 
the magnetisation current oscillates around zero and vanishes in average.
Within this DNLS formulation, the spin transfer torque that appears in Eq.(\ref{eq:continuity}) plays the role of a magnon chemical potential, that by controlling the lifetime of the excitations, controls also the SW current between them.

The diagonalization of the LLG equation in a confined geometry leads to a discrete series of normal modes having each a different 
eigen-value, $\omega/(2\pi)$, the so-called Larmor precession frequency. The normal modes of the system are numbered
according to the number of half waves in the vibration. In the case of a 2D axially symmetric structure, the normal modes are identified by
two integers: $\ell$ and $m$, respectively the mode number in the azimuthal and radial directions. The analytical expression of the
normal modes of a perpendicularly magnetized disk is found in Refs. [\onlinecite{damon61,naletov11}]

\be\label{bessel}
c_{\ell,m}(r,\phi,t)=J_\ell (k_{\ell,m} r) e^{+ i \ell \phi} e^{- i \omega_{\ell m} t},
\ee
where $J_\ell$ are the Bessel functions of the first kind and $k_{\ell,m}$ is the modulus of the in-plane SW wave-vector, which
depends on the boundary conditions.

The above labelling can be extended to the case of two different magnetic disks coupled by dipolar interaction. 
In the perpendicular geometry, the strength of the dynamical dipolar coupling is attractive (lower in energy) when both layers vibrate in antiphase, 
because the dynamical dipolar charges in each layer are alternate \cite{belmeguenai07,gubbiotti04}. 
Thus the binding state  $B$ corresponds to a collective motion where the two layers vibrate anti-symmetrically and the anti-binding state $A$ to a collective 
motion where the two layers vibrate symmetrically. The $B$ modes correspond to a precession amplitude that is larger in the thick layer $b$, while the $A$ modes
correspond to a precession amplitude that is larger in the thin layer $a$ \cite{naletov11}, see Fig.\ref{fig:figure1} for a cartoon.

The dynamical dipolar coupling does not modify the nature of the modes hence, in order to describe the dynamics of
the bi-layer system, we shall just add a new index $B$ or $A$ indicating if the precession occurs in antiphase (mostly in
the thick layer) or in phase (mostly in the thin layer), respectively. There are thus three indices to label the observed eigen-modes: the usual
azimuthal and radial indices for a single disk ($\ell,m$), plus an additional index referring to the symmetrical
or anti-symmetrical ($A$ or $B$) coupling between both layers.  

The identification of the SW modes and their symmetry is essential to couple the oscillator to an external source, since SW modes can couple only to a 
source with the same symmetry. Here the $\ell$ index determines the rotational symmetry of the SW mode. The $\ell=0$ modes correspond to SW that do not rotate in the $x$-$y$ plane and can be excited only by a spatially uniform in-plane RF field,
while the $\ell=1$ modes correspond to SW that rotate around the disk in the same direction as the Larmor precession, and can be excited only
by an RF Oersted field with orthoradial symmetry. Thus, exciting the system with these different means gives two different spectra. 

\section{Micromagnetic simulations at zero current} %

In this section we describe the SW spectra by means of micromagnetic simulations without spin-transfer torque.
Those simulations were performed with the NMAG micromagnetic software \cite{fischbacher07}, where the sample is described by finite element  tetrahedral mesh. The latter has a maximum intersite distance of 6 nm, 
of the order of the Py exchange length. The micromagnetic parameters are the same used in Ref.[\onlinecite{naletov11}] and are reported in Tab.I for convenience.
The dynamics at each of the $i=1,...,N$ nodes of the mesh of disk $j=(a,b)$ is described by the following LLG equation for the unit magnetisation vector, $\bm{m}_i^j=\bm{M}_i^j/M_s$:

\begin{table}\label{tab:torque_parameters}
\begin{center}
\begin{tabular}{|c|c|c|c|c|c|}
\hline
$4\pi M_a$ (G) & $\alpha_a$ & $4\pi M_b$ (G) & $\alpha_b$ & $\gamma$ (rad.s$^-1$.G)\\
\hline
$8.2\times 10^3$ & $1.5\times 10^{-2}$ & $9.6\times 10^3$ & $9\times 10^{-3}$ & $1.87 \times 10^7$\\
\hline
\end{tabular}
\end{center}\caption{Parameters of thin (a) and thick (b) layer used in micromagnetics simulations.}
\end{table}

\be\label{eq:nmagllg}
\dot{\bm{m}}_i^j=-\gamma(\bm{m}_i\times\bm{H}_{{\rm{eff}}i}^j)+\frac{\alpha}{M_s}(\bm{m}^j_i\times\dot{\bm{m}}_i^j)
\ee

The integration of the LLG equation at each mesh site is performed by the Sundials ODE solver \cite{hindmarsh05}, which is based on variable steps multistep methods. The field $\bm{H}^j_i$ at each mesh node has contributions from applied field, first neighbour exchange
interaction and long range dipolar interaction, responsible for the coupling between the layers. 
The quantity of interest is the space-averaged magnetisation $\average{{\bm{m}}^j(t)}=\frac{1}{V^j}\int_{V^j}{\bm{m}}^j(\bm{r}^j,t)d^3{\bm{r}}$, which for our finite-elements mesh reduces to
\be
\bm{m}^j=\frac{1}{N}\sum_{i=1}^N\bm{m}_i^j.
\ee

\begin{figure}
\includegraphics[keepaspectratio,width=0.7\linewidth]{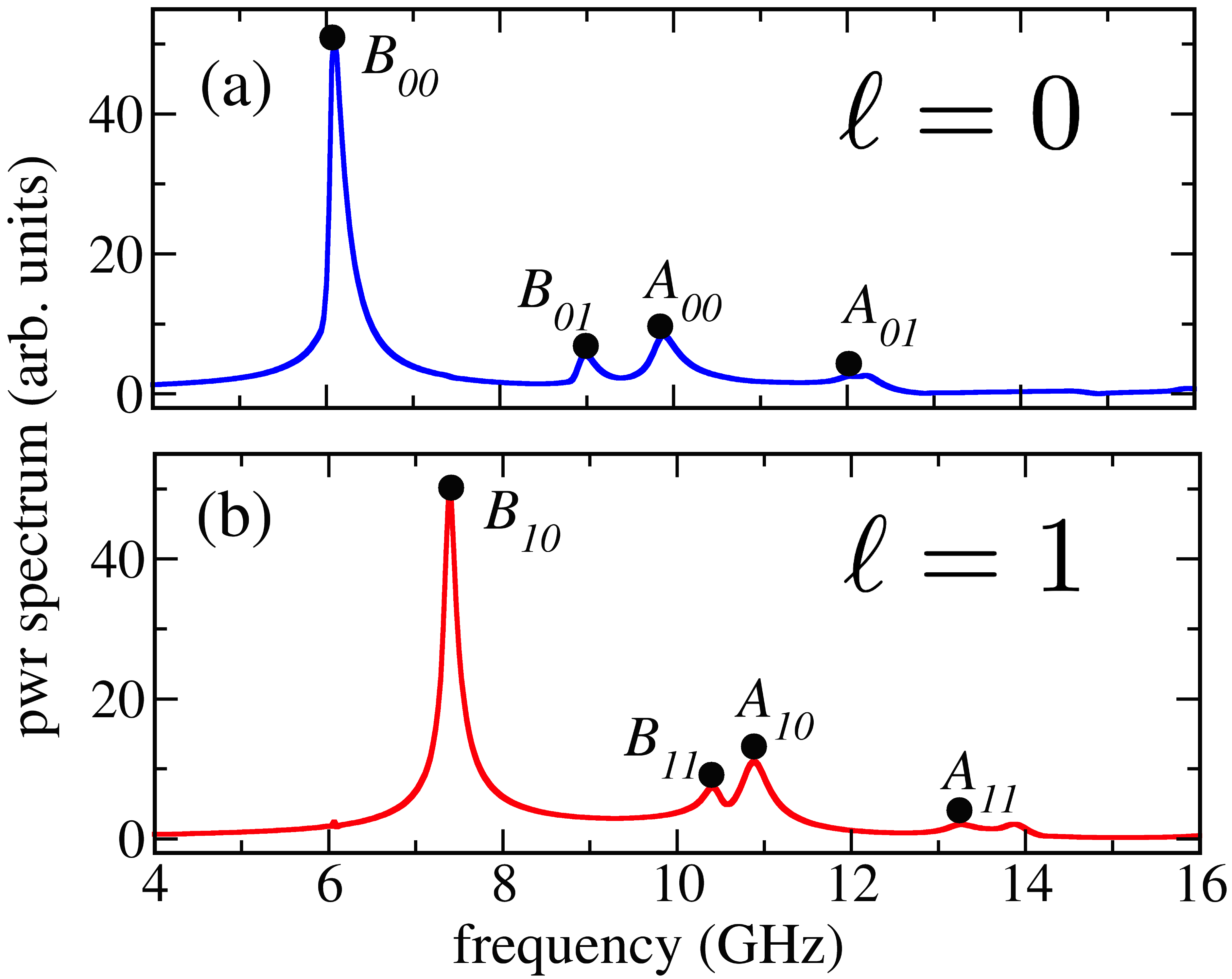}
\caption{(Color online) SW power spectrum for the $\ell=0$ (a) and $\ell=1$ (b) modes, obtained from the volume-averaged magnetisation. The $B$ (resp. $A$) modes corresponds to anti phase (resp. in phase) precession, 
whose amplitude is larger in the thick (resp. thin) layer.
\label{fig:figure2}}
\end{figure}

From this quantity, the SW amplitudes $c_j(t)$ are calculated.
The power spectrum, shown in Fig. \ref{fig:figure2}, is given by the Fourier transform of the time series of the collective SW amplitude averaged over the sample thicknesses: 
$c=(c_at_a+c_bt_b)/(t_a+t_b)$.

The modes with $\ell=0$ (displayed in blue tones) are excited starting from an initial condition where the magnetization uniformly tilted $8^\circ$ in the x direction with respect to the precession axis $\bm{z}$.
Instead, the the modes with $\ell=+1$ (displayed in red tones) are excited by applying to the magnetization aligned with the $\bm{z}$ axis the orthoradial vector field perturbation
$\theta(r,z)=\epsilon\hat{\bm{z}}\times\hat{\bm{\rho}}$. Here $\epsilon=0.01$ and $\hat{\bm{\rho}}$ is the unit vector in the radial direction.
Starting from these conditions, we have computed the time evolution of the system for 120 ns, with an integration time step of 5 ps. From the Fourier transform, the maximum frequency is 100 GHz and the frequency resolution 15 MHz.
The frequencies of the SW modes are displayed in Tab.\ref{tab:zeros} and \ref{tab:ones} and compared with the experimental values.

\begin{table}
\begin{center}
\begin{tabular}{|c|c|c|c|c|c|}
\hline
\hline
Exp.  $f$ (GHz) & 6.08 & 8.95 & 9.82 & 11.98 \\
\hline
Sim.   $f$ (GHz) & 6.08 & 8.94 & 9.83 & 12.00\\
\hline
SW modes      & $B_{00}$ & $B_{01}$ & $A_{00}$ & $A_{01}$ \\
\hline
\hline
\end{tabular}\caption{Comparative table of the resonance fields of the $\ell=0$ SW modes. Top are the peak locations measured experimentally in Ref.[\onlinecite{naletov11}]
Bottom are the eigen-frequencies extracted from the simulation with applied field $H_{\rm{ext}}=1$~T along $z$.}
\label{tab:zeros}
\end{center}

\begin{center}
\begin{tabular}{|c|c|c|c|c|c|}
\hline
\hline
Exp.  $f$ (GHz) & 7.44 & 10.47 & 10.85 & - \\
\hline
Sim.   $f$ (GHz) & 7.46 & 10.46 & 10.90 & 13.85\\
\hline
SW modes      & $B_{10}$ & $B_{11}$ & $A_{10}$ & $A_{11}$ \\
\hline
\hline
\end{tabular}\caption{Comparative table of the resonance fields of the $\ell=+1$ SW modes. Top are the peak locations measured experimentally in Ref.[\onlinecite{naletov11}]. The mode $A_{11}$ is not visible in experimental data.
Bottom are the eigen-frequencies extracted from the simulation with applied field $H_{\rm{ext}}=1$~T along $z$.}
\label{tab:ones}
\end{center}
\end{table}

\section{Continuous Random Matrix Theory (CRMT) for spin transport}\label{sec:crmt}%

This section contains a thorough review of the CRMT semi-classical theory of spin-dependent transport in
magnetic multilayers. We follow closely the material presented in Refs. [\onlinecite{waintal00,rychkov09,borlenghi11}] and we extend it by providing an explicit formula for spin torque, that will be used in 
micromagnetics simulations of next section.
 
\subsection{Scattering matrix approach}\label{subsec:crmt1}%

Within the scattering matrix formulation developed by Landauer and Buttiker \cite{buttiker85}, a sample is defined by the scattering matrix $S$ which expresses the outgoing propagating modes in term of the incoming ones.
The incoming modes are filled according to the Fermi-Dirac distribution of the leads to which they are connected. From the elements of the scattering matrix various physical quantities can be calculated, 
such as the conductance $G$, the spin currents $\bm J$ and charge current $I$.

\begin{figure}\label{fig:S1}
\includegraphics[keepaspectratio,width=0.7\linewidth]{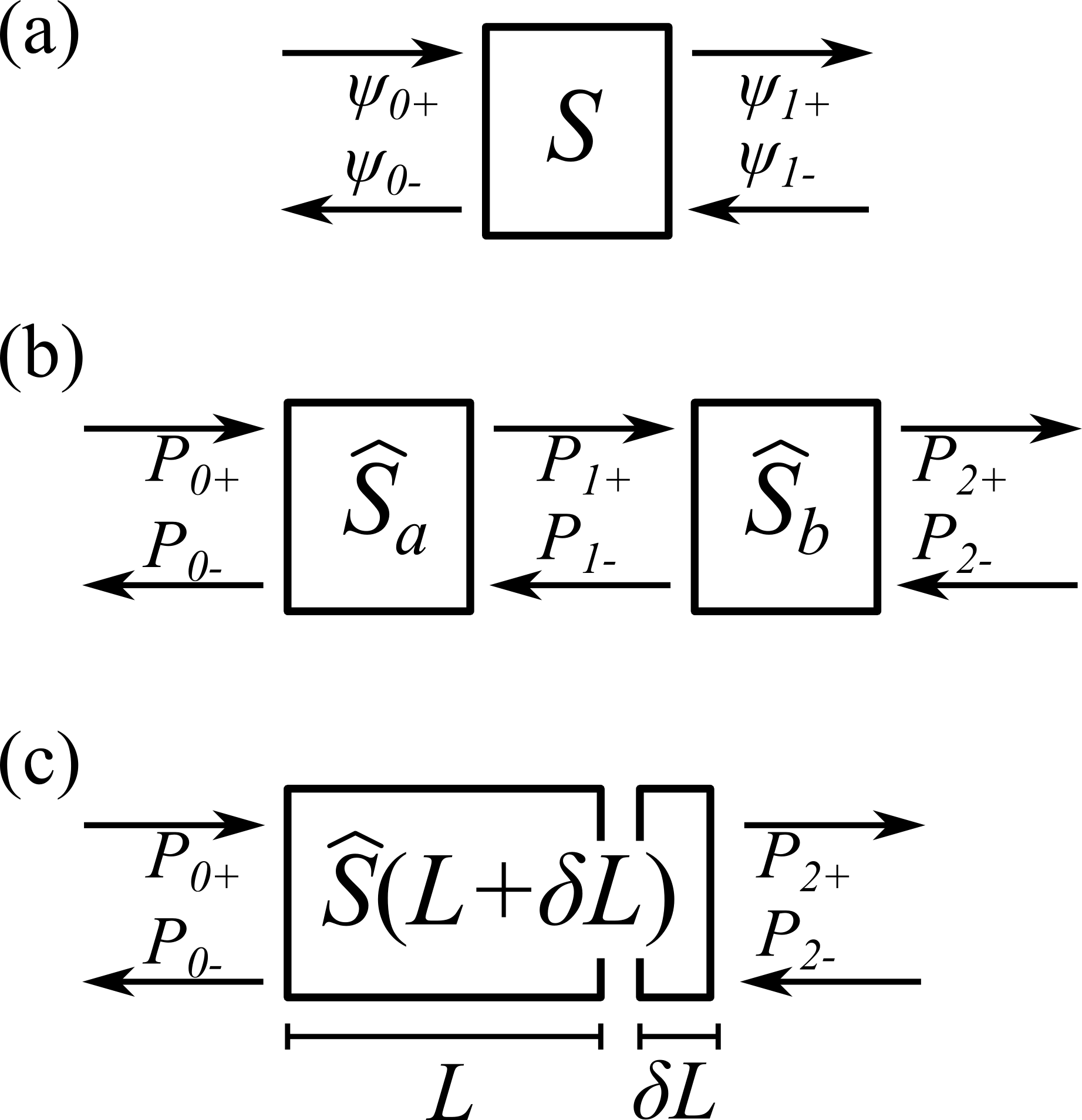}
\caption{(a) Scattering matrix relating incoming and outgoing modes $\psi_n$ in region $n=0,1$. The $+$ and $-$ sign represent respectively right and left propagating modes.
(b) Cartoon of the "hat" matrix approach. We define the region 0, 1 and 2 respectively on the left, middle and right of the two samples $S_a$ and $S_b$.
 In each region the 4-vectors ${\bm P}_{i\pm\sigma}$ represents the probability to find right ($+$) propagating and left ($-$) propagating electrons. 
 (c) The CRMT equations allow one to calculate the hat matrices as a function of the length $L$ of the sample, using the sum law for hat matrices.}
\end{figure}

The system contains $N_{ch}\gg 1$ propagating modes (or channels) per spin. In particular, one has $N_{ch}\approx A/\lambda_F^2$, where $A$ is the transverse area of the 
electrode and $\lambda_F$ is the Fermi wavelength. The amplitude of the wave-function on the different modes is given by the vector ${\bm \psi}_{i\pm}$ with $N_{ch}$ elements

\be
{\bm \psi}_{i\pm}=\left(\begin{array}{c}{\bm\psi}_{i\pm\ua} \\{\bm\psi}_{i\pm\da}\end{array}\right).
\ee
The latter contains the amplitudes for the right (+) and left (-) moving electron direction with spin 
$\sigma=\u,\d$ along the $z$-axis in region  $i=0,2$ of the multilayer, see Fig.\ref{fig:S1} for a cartoon.
The $S$ matrix is a $4N_{ch}\times4N_{ch}$ unitary matrix that relates the outgoing modes
to the ingoing ones: 

\be
\label{eq:scattering-formalism}
\left(\begin{array}{c}{\bm\psi}_{0-} \\{\bm\psi}_{1+}\end{array}\right)=S\left(\begin{array}{c}{\bm\psi}_{0+}\\{\bm\psi}_{1-}\end{array}\right), 
\ee
see Fig.\ref{fig:S1} (a) for a schematic of the system.

and consists of $2N_{ch}\times2N_{ch}$ transmission $t,t'$ and reflection $r,r'$ sub-blocks.
\be
\label{eq:s}
S=\left(\begin{array}{cc}  r' & t \\ t' & r \end{array}\right).
\ee	
Here the $(r,t)$ and $(r^\prime,t^\prime)$ describe reflection and transmission respectively from left to right and from right to left. 
The transmission and reflection matrices have an internal spin structure:

\be
t=\left(\begin{array}{cc}t_{\u\u} & t_{\u\d} \\ t_{\d\u} & t_{\d\d}\end{array}\right)
\ee
where $t_{\sigma\sigma'}$ are $N_{ch}\times N_{ch}$ matrices containing amplitudes for transmission between $\sigma'$ and $\sigma$ spin states, capturing both spin preserving and spin-flip phenomena.

The conductance of the system is given by the Landauer formula \cite{buttiker85}

\be\label{eq:landauer}
G=\frac{e^2}{h}\Tr \left[t^\dag t\right],
\ee 
while the spin current in region 0 reads

\be\label{eq:j}
\frac{\partial\vec J_0}{\partial\mu}=\frac{1}{4\pi}\Tr\left[t\vec\sigma t^\dag\right], 
\ee
$\mu$ being the difference of chemical potential between the two electrodes \cite{waintal00}. 

\subsection{Random Matrix Theory (RMT)}\label{subsec:crmt2}%

The scattering matrix approach is fully quantum and contains interference effects such as weak localization or universal conductance fluctuations \cite{waintal00,montambaux07}. 
The system studied in this paper is a nanopillar of 250 nm of diameter connected to top and bottom electrodes. Those contain $\approx 10^4-10^5$ propagative channels. Here
the scattering is not perfectly ballistic (mismatch at the interfaces, surface roughness or impurity scattering) so that channels get mixed up. Random Matrix Theory 
(RMT) \cite{beenakker97,mehta04,waintal00} assumes that this mixing is \emph{ergodic}: an electron entering the system in a given mode will leave it in an arbitrary mode, acquiring a random phase in the process.

In this case, the transmission and reflection probabilities of an electron are well caracterized by their average over the propagative channels.
This average is obtained by taking the trace over the $N_{ch}$ of the original reflection and transmission matrices. 
For instance, the hat matrix $\hat t$ \cite{rychkov09,borlenghi11} is defined as
\be
\label{eq:hat}
\hat t_{\sigma\eta,\sigma'\eta'}=\frac{1}{N_{ch}}\Tr_{N_{ch}}[t_{\sigma\sigma'} t^\dag_{\eta\eta'}].
\ee
Explicitely, this reads
\be
\label{eq:t-hat} 
\hat{t}=\frac{1}{N_{ch}} \Tr_{N_{ch}}
\roundb{
\begin{array}{cccc} t_{\uparrow \uparrow} t_{\uparrow \uparrow}^{\dagger} &
t_{\uparrow \uparrow}t_{\uparrow \downarrow}^{\dagger} &
 		    t_{\uparrow \downarrow}t_{\uparrow \uparrow}^{\dagger} &
t_{\uparrow \downarrow}t_{\uparrow \downarrow}^{\dagger} \\
 		    t_{\uparrow \uparrow}t_{\downarrow \uparrow}^{\dagger} &
t_{\uparrow \uparrow}t_{\downarrow \downarrow}^{\dagger} &
 		    t_{\uparrow \downarrow}t_{\downarrow \uparrow}^{\dagger} &
t_{\uparrow \downarrow}t_{\downarrow \downarrow}^{\dagger} \\
                    t_{\downarrow \uparrow}t_{\uparrow \uparrow}^{\dagger}
&  t_{\downarrow \uparrow}t_{\uparrow \downarrow}^{\dagger} &
 		    t_{\downarrow \downarrow}t_{\uparrow \uparrow}^{\dagger} &
t_{\downarrow \downarrow}t_{\uparrow \downarrow}^{\dagger} \\
                    t_{\downarrow \uparrow}t_{\downarrow
\uparrow}^{\dagger} &  t_{\downarrow \uparrow}t_{\downarrow
\downarrow}^{\dagger} &
 		    t_{\downarrow \downarrow}t_{\downarrow \uparrow}^{\dagger} &
t_{\downarrow \downarrow}t_{\downarrow \downarrow}^{\dagger}
\end{array}},
\ee
with the same structure for the reflection hat matrix.

The elements of this matrix correspond to the probability for an electron with a given spin to be transmitted ($\hat{t}$) or reflected ($\hat{r}$) by the system.
In particular, the terms $(1/N_{ch}) {\rm Tr_{N_{ch}}}( t_{\ua \ua}t_{\ua \ua}^{\dagger})\equiv T_{\ua \ua}$ and $(1/N_{ch}) {\rm Tr_{N_{ch}}}( t_{\da \da}t_{\da \da}^{\dagger})\equiv T_{\da \da}$ correspond to the probability to transmit an electron with up and down spin correspondingly. The terms at the corners, $(1/N_{ch}) {\rm Tr_{N_{ch}}}( t_{\ua \da}t_{\da \ua}^{\dagger})\equiv T_{\ua\da}$ and $T_{\da\ua}$ correspond to transmission probabilities with spin flip.
The so-called "mixing transmission", $(1/N_{ch}) {\rm Tr_{N_{ch}}}( t_{\ua \ua}t_{\da \da}^{\dagger})=T_{mx}$ is a complex number whose amplitude measures how much of a spin transverse to the magnetic layer can be transmitted through the system, while its phase amounts for the corresponding precession. $T_{mx}$ decays exponentially with the size of the ferromagnet, and accounts for spin transfer effect \cite{waintal00,borlenghi11}. 

The other off diagonal elements in the hat matrix Eq.(\ref{eq:hat}) can be ignored \cite{waintal00,borlenghi11}. In the basis parallel to the magnetisation, Eq.(\ref{eq:t-hat}) therefore becomes:
\be\label{eq:CRMT-t-hat}
\hat{t}=\roundb{
\begin{array}{cccc} T_{\ua\ua} &0 & 0& T_{\ua\da} \\
 		   0 & T_{mx} & 0 & 0 \\
                   0 &  0 & T_{mx}^* & 0 \\
                    T_{\da\ua} &  0&0 & T_{\da \da}\end{array}},
\ee

with the same structure for the reflection matrix $\hat{r}$. The hat-matrix $\hat S$ has a form similar to Eq.(\ref {eq:s}),

\be\label{eq:hat_s}
\hat S=
\left(\begin{array}{cc}  \hat r' & \hat t \\ \hat t' & \hat r \end{array}\right).
\ee	

In order to describe transport in non-collinear multilayers, where the orientation of the magnetization
changes inside the system, one needs to rotate the original $S$ matrix as $\tilde S= R_{\theta,\vec n}S R_{\theta,\vec n}^\dagger$ in the chosen working basis. Here the matrix 

\be\label{eq:rotate}
R_{\theta,\vec n}=\exp(-i\vec\sigma\cdot\vec n \ \theta/2)
\ee
is the rotation matrix of angle $\theta$ around the unit vector $\vec n$ that brings the magnetization onto the z-axis of the working basis. In term of hat matrices, this translates directly into 
\be\label{eq:hatrotate}
\hat{\tilde S}= \hat R_{\theta,\vec n}\hat S \hat R_{\theta,\vec n}^\dagger,
\ee

with $\hat R_{\sigma\eta,\sigma'\eta'}=R_{\sigma\sigma'} R^*_{\eta\eta'}$ a unitary matrix.

From Eq.(\ref{eq:CRMT-t-hat}), the conductance is given by

\be\label{eq:average_conductance}
G=\frac{1}{\mathcal{R}_{sh}}\roundb{T_{\ua\ua}+T_{\ua\da}+T_{\da\ua}+T_{\da\da}}
\ee
where the Sharvin resistance $\mathcal{R}_{sh}=\frac{h}{N_{ch}e^2}$, is a material property that can be experimentally measured, related to the number $N_{ch}$ of transverse propagative channels for the electrons
crossing the system. 
Eq.(\ref{eq:average_conductance}) is analogous to the Landauer formula Eq.(\ref{eq:landauer}) and consists of the sum of all the possible transmission processes (spin preserving and spin flipping) for an electron.

In analogy with the modes $\psi_{i\pm\sigma}$ for the scattering matrix, the $4$-vectors  ${\bm P}_{i\pm}$ are introduced 
\be 
\label{eq:P_vector}
{\bm P}_{i\pm}=\roundb{\begin{array}{l}{P}_{i\pm,\u}\\{P}_{i\pm, mx}\\{P}^*_{i\pm, mx}\\{P}_{i\pm,\d}\\\end{array}}.
\ee
The components o${\bm P}_{i\pm \ua,\da}$  have interpretation in term of probabilities for an electron to propagate in the region $i$ of the system. 
The "mixing" components, ${P}_{mx}$ are complex numbers which correspond to probability to find the electron with spin along the $x$ (real part) or $y$ (imaginary part) axis. Inside magnetic layers where the $z$ axis xill correspond to the direction of the magnetisation, they will correspond to the probability for the spin to have a part transverse to the magnetisation.

For the following discussion, it is convenient to consider a system made of two conductors connected in series, 
described by the two hat matrices $\hat{S}_a$ and $\hat{S}_b$, see Fig. \ref{fig:S1} (b). The space is thus divided into three regions: region 0 and 2 respectively at the leftmost and rightmost part of the system, and region 1 in between the two hat matrices.
In analogy with Eq.(\ref{eq:s}) which expresses the amplitudes of the outgoing modes in term of the incoming ones, in subsystems $a$ and $b$ one has respectively
\beA
\label{eq:hat_scattering1}
\left(\begin{array}{c}{\bm P}_{0-}\\{\bm P}_{1+}\end{array}\right) &=& \hat S_{a}\left(\begin{array}{c}{\bm P}_{0+}\\{\bm P}_{1-}\end{array}\right),\\
\left(\begin{array}{c}{\bm P}_{1-}\\{\bm P}_{2+}\end{array}\right) &=& \hat S_{b}\left(\begin{array}{c}{\bm P}_{1+}\\{\bm P}_{2-}\end{array}\right),\\
\eeA
while for the total system $a+b$

\be
\label{eq:hat_scattering2}
\left(\begin{array}{c}{\bm P}_{0-}\\{\bm P}_{2+}\end{array}\right)=\hat S_{a+b}\left(\begin{array}{c}{\bm P}_{0+}\\{\bm P}_{2-}\end{array}\right).
\ee
Here we have used the \emph{addition law} of hat matrices, a fundamental property that will be useful in the derivation of spin currents and spin torque inside the system.
According to this rule, given the hat matrices for separate systems $a$ and $b$, the hat matrices of the composed system $a+b$ read \cite{rychkov09,borlenghi11}

\beA
\hat{t}_{a+b} &=& \hat{t}_{a} \frac{1}{\hat{1}-\hat{r}^\prime_{b}\hat{r}_{a}}\hat{t}_b\label{eq:tsum}\\
\hat{r}_{a+b} &=& \hat{r}_{b}+\hat{t}^\prime_{b}\frac{1}{\hat{1}-\hat{r}_{a}\hat{r}^\prime_{b}}\hat{r}_a\hat{t}_b\label{eq:rsum}
\eeA
with $\hat{1}$ the $4\times 4$ identity matrix. Similar expression holds for $r_{a+b}^\prime$ and $t_{a+b}^\prime$.

The main result of Refs.[\onlinecite{rychkov09,borlenghi11}] is that the spin current in region $i$ of the system can be expressed in terms of probability vectors as follows:
\be\label{spin_curr1}
\bm{J}_i=\frac{N_{ch}}{4\pi}{\vec{\bm\sigma}}\cdot (P_{i+}- P_{i-}), 
\ee
where $\vec{\bm{\sigma}}=(\vec{\sigma}_{\ua\ua},\vec{\sigma}_{\ua\da},\vec{\sigma}_{\da\ua},\vec{\sigma}_{\da\da})$ is the vector of components of Pauli matrices.
To express the spin current in region 1 in between the two conductors as a function of the hat matrices, we use Eqs.(\ref{eq:hat_scattering1})-(\ref{eq:rsum}) to eliminate $P_{2+}$ and $P_{2-}$ and obtain

\be
\label{eq:hatsum}
\left(\begin{array}{c}{\bm P}_{1+}\\{\bm P}_{1-}\end{array}\right)=
\left(\begin{array}{cc}  \frac{1}{\hat{1}-\hat{r}_a\hat{r}_b^\prime}\hat {t}_a^\prime & \frac{1}{\hat{1}-\hat{r}_a\hat{r}_b^\prime}\hat{r}_a\hat {t}_b \\ 
                                   \frac{1}{\hat{1}-\hat{r}_b^\prime\hat{r}_a}\hat{r}_b^\prime \hat {t}_a^\prime & \frac{1}{\hat{1}-\hat{r}_b^\prime\hat{r}_a}\hat {t}_b\end{array}\right)
\left(\begin{array}{c}{\bm P}_{0+}\\{\bm P}_{2-}\end{array}\right)
\ee	
The theory is completed imposing boundary conditions on the incoming electrons on both sides of the system.
For normal electrodes one has,

\beA
\label{eq:boundary_conditions}
{\bm P}_{0+}=\left(\begin{array}{l}\mu_0\\0\\0\\ \mu_0\end{array}\right),\ 
{\bm P}_{2-}=\left(\begin{array}{l}\mu_2\\0\\0\\ \mu_2\end{array}\right)
\eeA
where $\mu_0$ and $\mu_2$ are the respective chemical potentials of the two electrodes.
The generalization to magnetic electrodes is done imposing different chemical potentials
for majority and minority electrons in the leads.
By taking as boundary conditions $\mu_0=eU$, with $U$ the potential difference between the two sides and $\mu_2=0$, The spin current finally reads

\be\label{eq:spin_curr2}
\mathcal{J}_1=\frac{N_{ch}}{4\pi}\bm{J}eU
\ee
where the spin current per channel and per unit of potential difference reads
\be\label{eq:J}
\bm{J}=\vec{\bm{\sigma}}\cdot\frac{1-\hat{r}_b^\prime}{\hat{1}-\hat{r}_a\hat{r}_b^\prime}\hat{t}_a. 
\ee

Eq.(\ref{eq:spin_curr2}) allows one to compute the spin current in the region between two bulk materials of arbitrary thicknesses. In the multilayer considered here, the currents $\bm{J}_{a/b\pm\delta}$ are
calculated at positions $\delta=\pm1$ nm before and after the two magnetic layers, as shown in Fig.\ref{fig:figure4}. The  torque is the spin current absorbed by each layer, i.e. the quantity
\be\label{eq:tau}
\bm{\tau}^j = \frac{N_{ch}}{4\pi}f_jeU
\ee
with $f_j=\bm{J}_{j-\delta}-\bm{J}_{j+\delta}$, $j=a,b$.

\subsection{From scattering matrices to CRMT}\label{subsec:crmt3}%

In order to calculate the current in different regions of the multilayer, one needs to calculate the matrices $\hat{S}(L)$ as a function of the position $L$ inside the system. 
The main result of Refs.[\onlinecite{rychkov09,borlenghi11}] is that the matrix $\hat S(L+\delta L)$ for an infinitesimal increment of the position $\delta L$ is entirely characterised by
two matrices $\Lambda^t$ and $\Lambda^r$, defined as
\be\label{eq:thin}
\hat t(\delta L) = 1 - \Lambda^t \delta L  \ \ , \ \ \hat r(\delta L) = \Lambda^r \delta L
\ee
Once $\hat S(\delta L)$ is known, one can make use of the addition law Eqs.(\ref{eq:tsum}) and (\ref{eq:rsum})
to obtain a differential equation to compute $\hat S(L+\delta L)$. By taking the limit $\delta L\rightarrow 0$ one gets  the two
CRMT differential equations that describe hat matrices as a function of the length of the system:   

\beA
\label{eq:drdl}
\frac{\partial\hat r}{\partial L} &=& \Lambda^r -\Lambda^t\hat r -\hat r\Lambda^t 
+\hat r\Lambda^r\hat r \\
\label{eq:dtdl}
\frac{\partial\hat t}{\partial L} &=& -\Lambda^t\hat t + \hat r\Lambda^r t
\eeA

A bulk magnetic material is characterised by four independent parameters $\Gamma_\uparrow$, $\Gamma_\downarrow$ $\Gamma_{\rm sf}$ and $\Gamma_{\rm mx}$, which describe spin preserving and spin flip phenomena. 
With this parametrisation, one has for the transmission

\beA\label{eq:lambdaS}
\Lambda^t &=&
\roundb{\begin{array}{cccc}  
\Gamma_\ua +\Gamma_{\rm sf} & 0              &  0             & -\Gamma_{\rm sf} \\ 
           0                     & \Gamma_{\rm mx}&  0             & 0 \\
           0                     & 0              &\Gamma^*_{\rm mx} & 0 \\
        -\Gamma_{\rm sf}         & 0              & 0              & \Gamma_\da+\Gamma_{\rm sf} \end{array}},\\
\eeA
and for the reflection

\beA\label{eq:lambdaR}
\Lambda^r &=&
\roundb{\begin{array}{cccc}  
\Gamma_\ua -\Gamma_{\rm sf} & 0              &  0             & \Gamma_{\rm sf} \\ 
           0                     & 0 &  0             & 0 \\
           0                     & 0              &0 & 0 \\
        \Gamma_{\rm sf}         & 0              & 0              & \Gamma_\da-\Gamma_{\rm sf} \end{array}},
\eeA 
These four parameters correspond in turn to 5 different lengths. 
The two most important one are the mean free paths for majority ($\sigma=\ua$) and minority ($\sigma=\da$) electrons defined as $\ell_\sigma = 1/\Gamma_\sigma$. 
Next comes the spin diffusion length $l_{\rm{sf}} = [4\Gamma_{\rm{sf}} (\Gamma_\ua + \Gamma_\da)]-1/2$. 
Last come the complex number $\Gamma_{\rm{mx}} = 1/l_{\perp} + i/l_L$ where $l_\perp$ is the penetration length of transverse spin current inside the magnet while 
$l_L$ is the Larmor precession length. Upon integrating Eq. (\ref{eq:dtdl}) one obtains for the mixing transmission.

\be\label{eq:tmix}
T_{\rm{mx}}(L)=e^{-L/l_{\perp} -iL/{l_{\rm L}}}.
\ee
which shows the exponential decay of the transverse spin current, absorbed by the layer, giving the phenomenon of spin transfer torque.
The two lengths $l_\perp$ and $l_L$ are of the order of the nm \cite{borlenghi11}, since spin transfer torque is a phenomenon that occurs close to the interface.
For bulk materials, another important quantity is the current polarisation, $P_\sigma=\frac{T_{\ua\sigma}-T_{\da\sigma}}{T_{\ua\sigma}+T_{\da\sigma}}$. The latter decays exponentially as a function
of the distance in the material, with spin-flip length scale of the order of 15 nm for Py, and reaches a constant value. The behaviour of spin and charge currents in a multilayer for different magnetic configurations have been extensively studied in 
Ref.[\onlinecite{borlenghi11}], to which we refer for a thorough discussion.

The CRMT parameters ($\Gamma_\u$, $\Gamma_\d$
and $\Gamma_{sf}$) are in 1-1 correspondence with the Valet-Fert (VF) parameters 
\cite{valet93} [$\rho_\u$, $\rho_\d$ 
(resistivities for majority and minority electrons) and $l_{sf}$ (spin-flip diffusion length)]. 
Using the standard notations for the average resistivity $\rho^*$ 
and polarization $\beta$ [$\rho_{\uparrow(\downarrow)}= 2\rho^* (1\mp \beta)]$ one 
has \cite{rychkov09}:

\beA
\label{eq:C-RMT-VF}
\frac{1}{l_{\rm sf}} &=& 2\sqrt{\Gamma_{\rm sf}}\sqrt{\Gamma_\uparrow + \Gamma_\downarrow}\\
\beta&=&\frac{\Gamma_\downarrow - \Gamma_\uparrow}{\Gamma_\uparrow + \Gamma_\downarrow}\\
\frac{\rho^*}{{\cal R}_{\rm sh}}&=&(\Gamma_\uparrow+\Gamma_\downarrow)/4
\eeA
This parametrization does not fix the mixing coefficients $\Gamma_{mx}$, which
only play a role in non collinear configurations \cite{waintal00}. 

In the VF theory, the interfaces are characterised by the average resistance $r^{b*}$ and polarisation $\gamma$
with  $r_{\ua,\da} =2r^{b*}(1 \pm \gamma)]$. In the CRMT formalism, interfaces are modelled as a virtual material
with their own transmission and reflection hat matrices. Those are related to the VF parameters as follows 
\beA\label{eq:tint}
T_{\ua\ua} &=&\frac{(1+e^{-\delta})/2}{1+2(r^{b*}/\mathcal{R}_{sh})(1-\gamma)}\nonumber\\
T_{\da\ua} &=&\frac{(1-e^{-\delta})/2}{1+2(r^{b*}/\mathcal{R}_{sh})(1-\gamma)}\nonumber\\
T_{\ua\da} &=&\frac{(1+e^{-\delta})/2}{1+2(r^{b*}/\mathcal{R}_{sh})(1+\gamma)}\nonumber\\
T_{\da\da} &=&\frac{(1+e^{-\delta})/2}{1+2(r^{b*}/\mathcal{R}_{sh})(1+\gamma)}
\eeA

while for the reflection coefficient one has
\beA\label{eq:rint}
R_{\ua\ua} &=& 1-\frac{1}{1+2(r^{b*}/\mathcal{R}_{sh})(1-\gamma)}\nonumber\\
R_{\da\da} &=& 1-\frac{1}{1+2(r^{b*}/\mathcal{R}_{sh})(1+\gamma)}
\eeA
and $R_{\ua\da}=R_{\da\ua}=0$.

\begin{figure}
\begin{center}
\includegraphics[width=8.0cm]{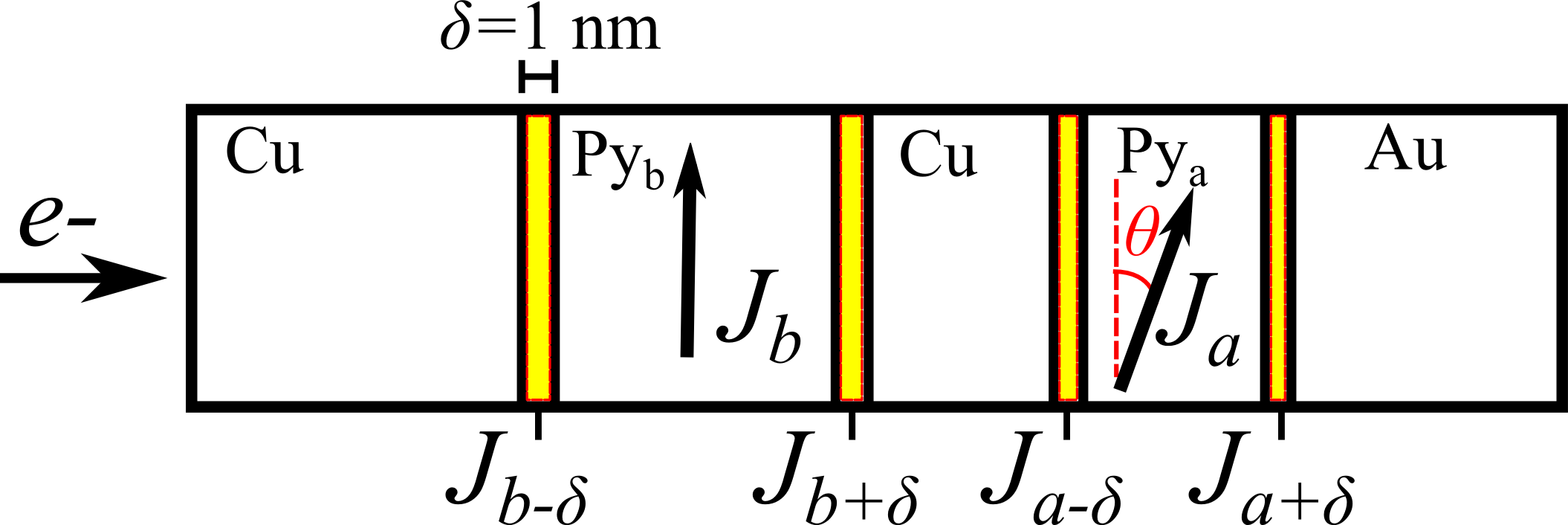}
\caption{(Color online) Cartoon of the nano-pillar simulated with CRMT. Electrons enter from the left and are spin-polarised along
the direction of $M_b$. When they enter the thin layer, they change their polarisation along the direction of $M_a$. Because of the
conservation of angular momentum, the transverse component of the spin-polarisation is transferred to the thin and thick layers as a spin torque} 
\label{fig:figure4}
\end{center}
\end{figure}

At this point the parametrisation of CRMT is complete. Eqs.(\ref{eq:drdl}) and (\ref{eq:dtdl}) have been solved numerically to simulate a one dimensional magnetic multilayer.
Each material and interface are simulated separately, and the whole system is recovered by applying the addition law Eqs.(\ref{eq:tsum}) and (\ref{eq:rsum}). 
 
 In order to calculate magnetoresistance and spin torque as a function of the magnetic configuration, we simulate transport the nano-pillar depicted in Fig.\ref{fig:figure4} by keeping the magnetisation $M_b$ fixed 
 and rotating the magnetisation $M_a$ of an angle $\theta$ in spin space. Such rotation is obtained by applying to the hat matrices of the thin layer the transformation Eq.(\ref{eq:hatrotate}). 
 


\begin{figure}
\begin{center}
\includegraphics[width=7.0cm]{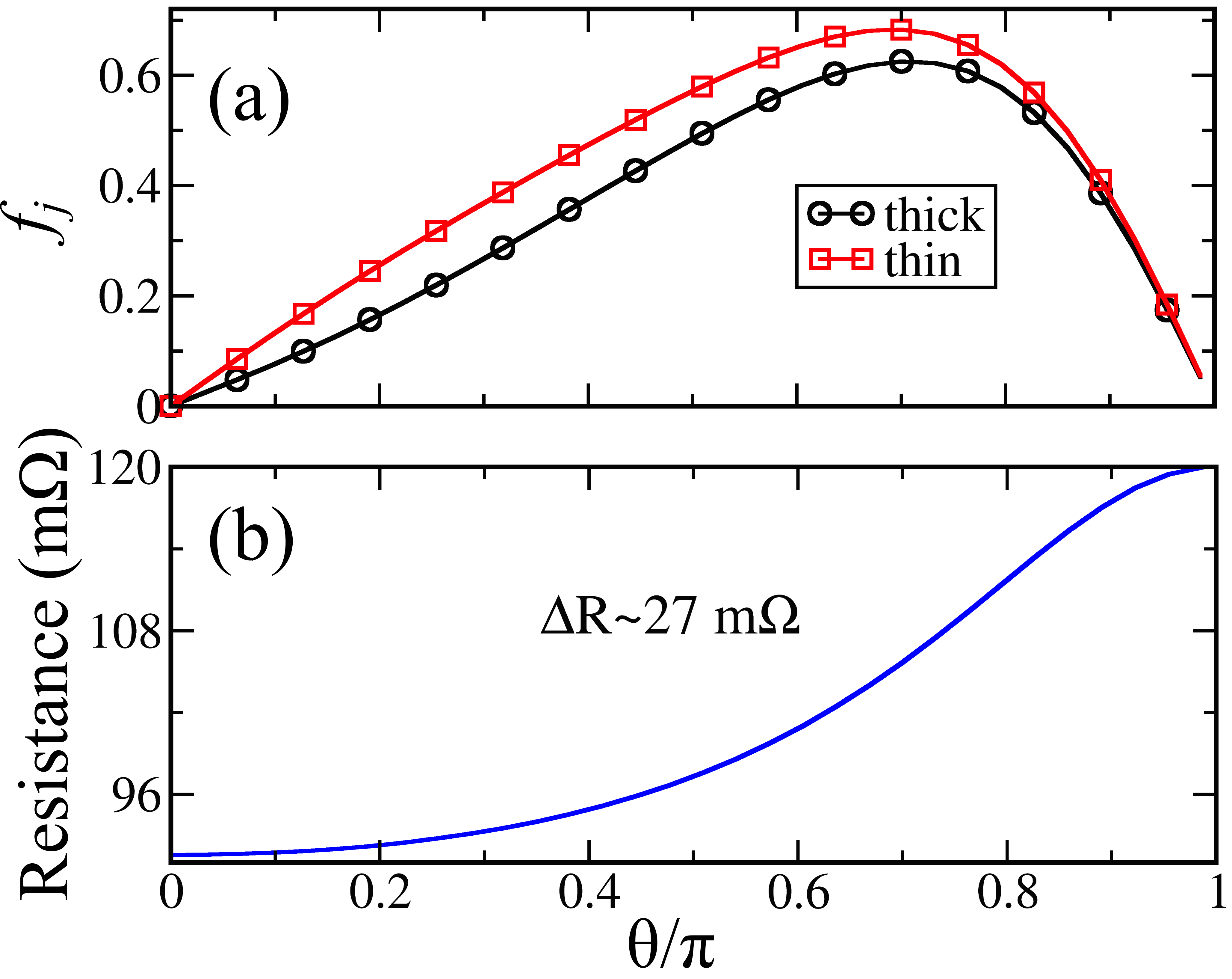}
\caption{(Color online) Calculation of the angular dependence of spin torque (a) and resistance (b) obtained keeping $\bm{m}_b$ fixed and rotating $\bm{m}_a$.
Magneto-resistance hysteresis curve of the nanopillar for an in-plane magnetic field. The dark (respectively light) symbols indicate the magnetic field being ramped up (respectively down)} 
\label{fig:crmt}
\end{center}
\end{figure}

Fig.\ref{fig:crmt}a) and b) shows respectively the angular dependence of spin torque and resistance inside the nanopillar for a single propagative channel and with $eU=1$. 
The resistance difference between parallel and antiparallel configuration is 27 m$\Omega$. We remark that this calculation, performed without adjustable parameters, reproduces 
experimental data within the $10\%$.

We note also that in our system, the current is spin-polarized by the thick layer and impinges the thin layer, exerting a torque 
that destabilizes its magnetization. Because of multiple reflections of spin polarized electrons between the two Py layers, 
STT tends to stabilize the thick layer, increasing its effective damping. Thus, at low current, one can consider the thick layer as "fixed" and the thin layer as "free''. However, at high current both layers undergo a coupled precession (see next section).

\section{Simulations of current driven dynamics}%
 
This section contains the main results of the paper. Here solve simultaneously the transport equations coupled to the LLG equation with spin transfer torque, by coupling CRMT to Nmag.
The approach described here is valid for one dimensional systems, where the magnetisation varies only along the direction of propagation of electrons. In the present case the magnetisation is uniform along $z$
inside the material, but its dynamics is different in each layer.
A generalization of CRMT for a fully three dimensional spin transport has been recently developed \cite{petitjean12}. However, for the case considered here the one 
dimensional CRMT approach captures the physics well and is extremely fast. The other advantage of CRMT is that it is parametrised by the 
same set of experimentally accessible parameters as the VF theory (reported in Tabs.\ref{Tbulk} and \ref{Tint}), so that no free parameter is needed to characterise realistic systems and materials.

\begin{table}
\begin{center}
\begin {tabular}{|c|c|c|c|}
\hline
Material & $\rho*$ & $\beta$ & $1/l_{\mbox{sf}}$ \\
\hline
Cu & 5 & 0 & 0.002 \\
\hline
Au & 20 & 0 & 0.033 \\
\hline
Py & 291 & 0.76 & 0.182 \\
\hline
\end {tabular}
\end{center}
\caption{Bulk VF parameters for the materials which constitute the multilayer. The VF bulk resistivity $\rho^*$ is expressed in units $10^{-9}\Omega m$ and the spin-flip $l_{\mbox{sf}}$ length in $nm$.}
\label{Tbulk}

\begin{center}
\begin {tabular}
{|c|c|c|}
\hline
Interface & $r_b^*$ & $\gamma$\\
\hline
Cu$|$Py & 0.5 & 0.7 \\
\hline
Au$|$Py & 0.5 & 0.77 \\
\hline
\end {tabular}
\end{center}
\caption{VF parameters for the interfaces between magnetic and non-magnetic material in our system.
The VF interface resistivity is expressed in units of $10^{-15}\Omega \mbox{m}^2$}
\label{Tint}
\end{table}

To include the effect of spin transfer torque into Nmag, the LLG equation Eq.(\ref{eq:nmagllg}) at node $i$ of disk $j=a,b$
needs to be modified as follows
\beA\label{eq:llg2}
\dot{{\bm m}}^j_i &=&-\gamma({\bm{m}^j_i}\times{\bm H}^j_{{\rm eff}i})-\frac{\alpha^j}{M_s^j}\left({\bm{m}^j_i}\times{\dot{\bm{m}}^j_i}\right)\nonumber\\
			  &+&\frac{g\mu_B}{\hbar C M_s}\bm{\tau}^j_i\hat{\bm w}_i^j. 
\eeA
Here $g$ is the Land{\'e} factor, $\mu_B$ is the Bohr magneton and $C$ is the volume associated to each site of the mesh. $\bm{\tau}^j_i$ is the torque given by Eq.(\ref{eq:tau})
For our relatively homogeneous mesh, we have taken this volume as the total volume $V_j$ of disk $j=a,b$  divided by the total number of sites $N_j$.
For simplicity, we have defined the vector $\hat{\bm{w}}_i^j=\bm{m}_i^j\times (\bm{m}_i^j \times \bm{m}_i^{j^{\prime}})$.

To include STT in our micromagnetic simulations, we adopt the following self-consistent loop:
\begin{enumerate}
\item{At time $t$ and site $i$ in disk $j$, Nmag computes the vector $\bm{m}_i^j$} 
\item{The magnetic configuration is passed to the CRMT solver which computes the torque $\bm{\tau}^j_i$}
\item{Then the quantity $\bm{m}_i^j (t)+{\bm{\tau}_i^j}(t)\Delta t$ is set as new initial condition for the Nmag solver, 
which performs the time integration of the LLG equation between times $t$ and $t+\Delta t$.}
\end{enumerate}
At this point spin torque is re-calculated as a function of the new magnetic configuration and the loop starts again. The integration
time step $\Delta t$ is of the order of the ps. 

\begin{figure}
\begin{center}
\includegraphics[width=9.0cm]{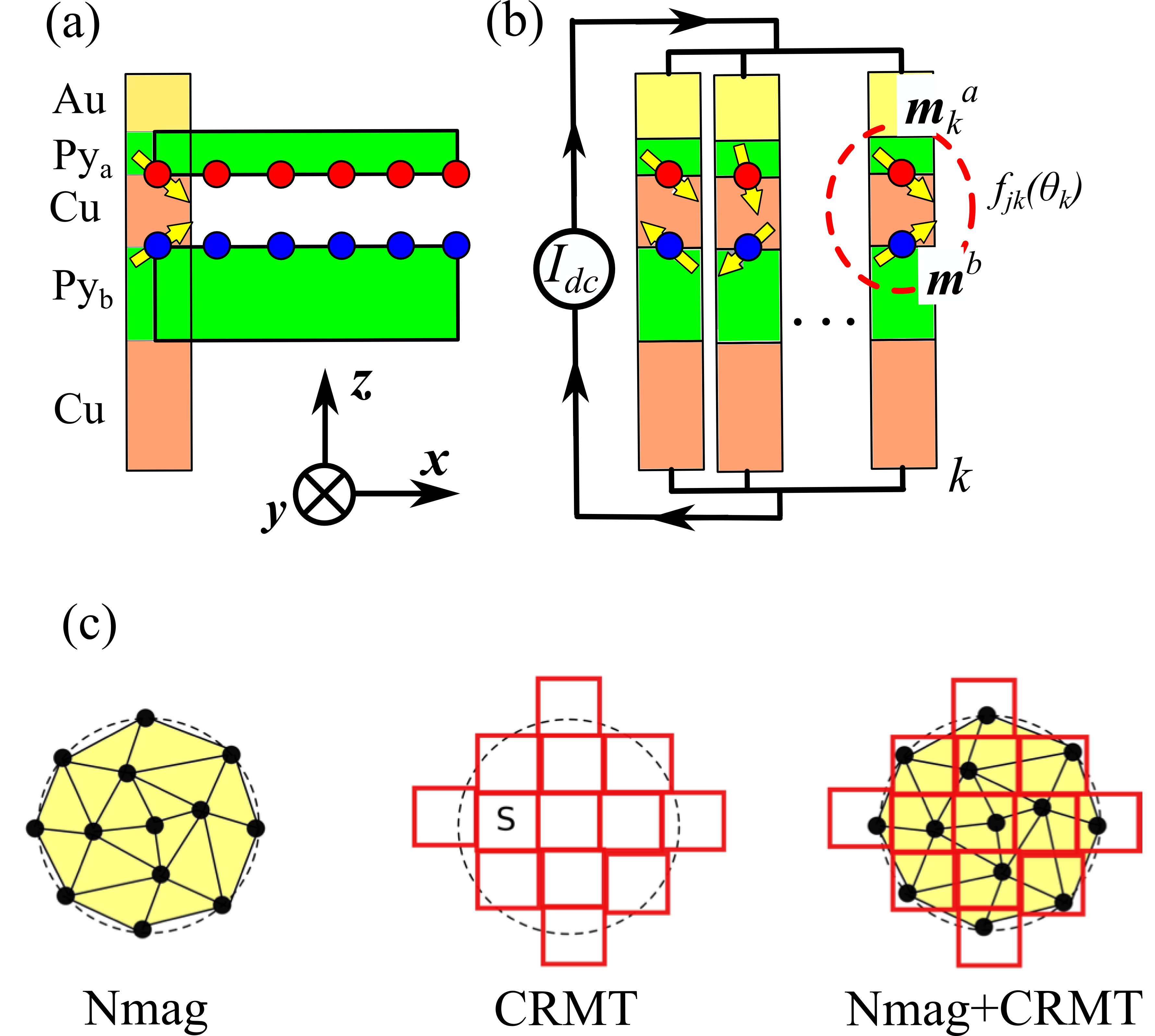}
\caption{Schematic of the method adopted to couple CRMT to Nmag, with the nano-pillar viewed from profile and the current flowing along
the $\bm{z}$ axis. a) One considers the sites lying on the inner surfaces of disks $a$ and $b$ that face each other. Then one selects a site of surface $1$ (thin layer, red dots)
and looks for its nearest neighbour among the sites of surface $2$ (thick layer, blue dots). b) To each couple of such sites, one associates a column $k$, that corresponds to a CRMT system,
with magnetisation ${\bm m}_1^k$ and ${\bm m}_2^k$. The system is then represented as an assembly of CRMT columns connected in parallel. From the magnetic configuration in each column $k$,
STT and conductance are computed.(c) Cartoon of the system divided into CRMT pillars, view from top.} 
\label{fig:columns}
\end{center}
\end{figure}

Since Nmag uses a finite element mesh while CRMT a finite difference discretisation of space, to implement the coupling, 
we divide the mesh into columns of sites, each column $k$ representing a CRMT 
pillar with cross section $S_k$. The whole system is then considered as an assembly of CRMT pillars connected in parallel, as shown in Fig.\ref{fig:columns}. 

Each column contains two sites, lying on 
surfaces of thin ($a$) and thick ($b$) layer facing each other, with magnetizations $\bm{m}_{ak}(t)$, and $\bm{m}_{bk}(t)$ correspondingly. 
From the angle $\theta_k$ between these two sites, one obtains the torques $\bm{\tau}^j_k(\theta)$ that act on the magnetisation vectors $\bm{m}^j_k$, and the resistance $R_k(\theta_k)$. 

This procedure takes into account the three dimensional texture of the magnetization, while electronic transport is along ${\bm z}$ only, without considering the lateral diffusion of spins.



Since the system is an assembly of columns connected in parallel, the current $I_k$ flowing in each column is given by the total current $I_{\rm{dc}}$
divided by the number of columns $N_k$, which in our system corresponds to the number of sites lying at the surface of each disk.
Each column has a cross section $S_k\approx S/N_k$. The current $I_k$  is related to the potential difference between the conductors via Ohm's 
law: $eU=R_k(\theta)I_{dc}/N_k$, where the resistance $R_k$ depends on the transmission probability $T_k(\theta)$ for an electron to cross column $k$:

\be\label{eq:sharvin_conductance}
R_{K}=\frac{\mathcal{R}_{sh}}{T_{k}(\theta)S_{k}}.
\ee
Inserting Eq.(\ref{eq:sharvin_conductance}) into Eq.(\ref{eq:torque_nmag1}), and recalling that $\mathcal{R}_{\rm sh}=h/(e^2N_{ch})$, the torque finally reads

\be\label{eq:torque_nmag2}
\bm{\tau}_{jk}=\frac{g\mu_B}{2eV_j}\frac{N_j}{N_k}I_{dc}\frac{f_{jk}(\theta)}{T_k(\theta)}\hat{\bm{w}}_j.
\ee
We remark that Eq.(\ref{eq:torque_nmag2}) is quite general and can be applied to systems with arbitrary geometry, provided
that electronic transport is one dimensional only.



\section{Results and discussion}%

\begin{figure}
\begin{center}
\includegraphics[width=9.0cm]{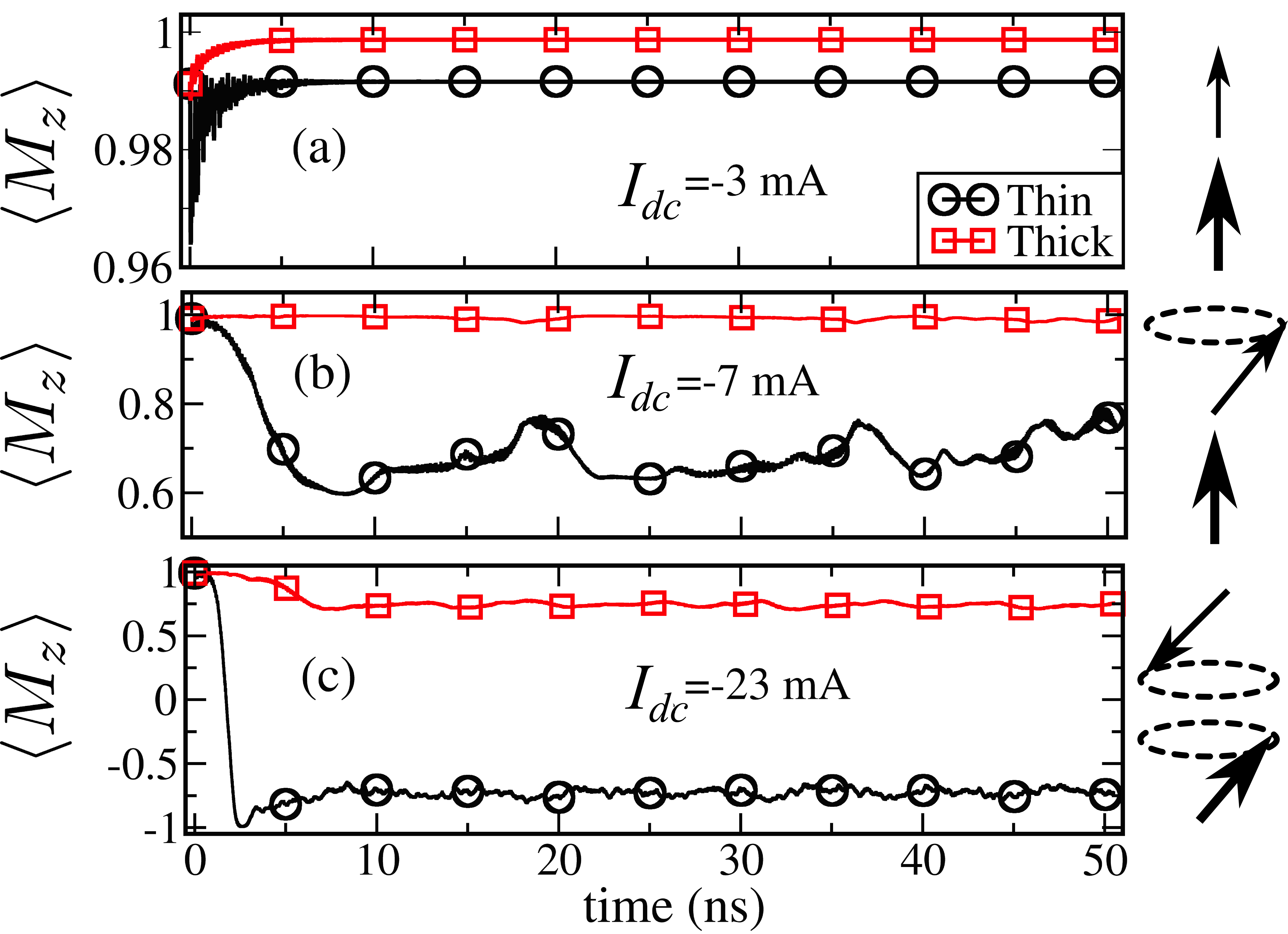}
\caption{Time evolution of $M_z$ for different values of the dc current. a) Subcritical regime, where both magnetisation vectors are aligned. b) Slightly higher than critical current, where
the thin layer precesses and the thick layer stand still. c) High current, with reversal of thin layer and coupled precession.} 
\label{fig:result1}
\end{center}
\end{figure}

We turn now to the discussion of the coupled Nmag-CRMT simulations.
A qualitative description of the different dynamical regimes is given by the time evolution of $M_z$, displayed in Fig.\ref{fig:result1}. Panel (a) shows the region beyond critical current, where both the magnetisations
are fixed  and aligned with the $\bm{z}$ axis. Panels (b) shows the region slightly beyond critical current, where the thin layer precesses and the thick one remains fixed. Finally, panel (c) displays the high current regime, where the thin layer is reversed
and both layers undergo coupled precession. This behaviour is due to the repulsive character of dipolar interaction: as the magnetisation of the thin layer reverses, it repels the magnetisation of the thick one, causing it to precess.

\begin{figure}
\begin{center}
\includegraphics[width=8.0cm]{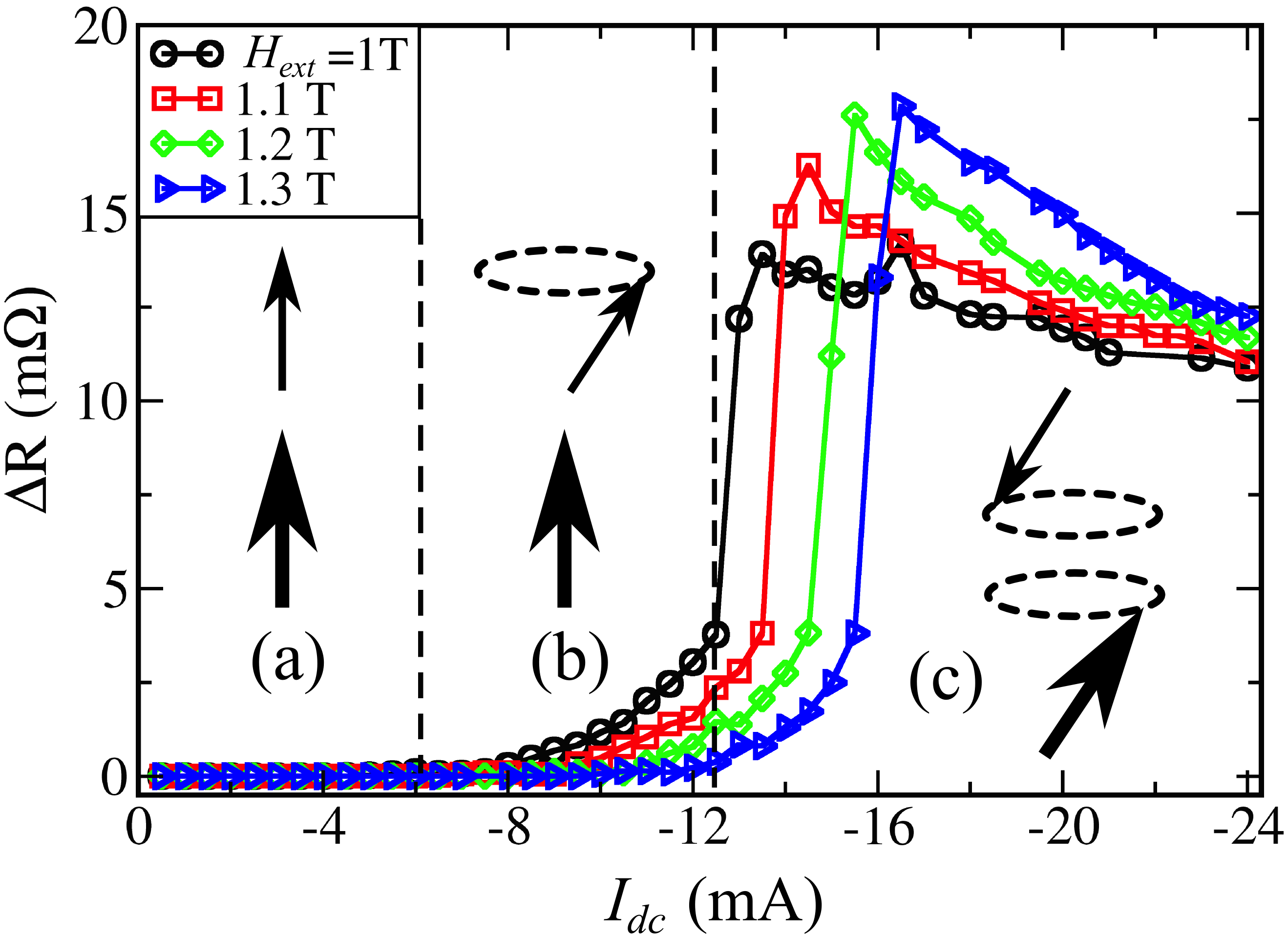}
\caption{Resistance difference $\Delta R$ between parallel and anti-parallel configuration as a function of the dc current $I_{dc}$, calculated for different values of the applied field $H_{\rm{ext}}$.} 
\label{fig:result2}
\end{center}
\end{figure}

Fig.\ref{fig:result2} shows the resistance as a function of the dc current, for values of the applied field between $1$T ands $1.3$T. One can recognise clearly the three dynamical regions discussed above. In particular, the resistance starts increasing around $-6$ mA,
indicating the region of critical currents. The resistance increases with the current until it reaches a peak between
$-13$ and $-16$ mA, corresponding to the reversal of the magnetisation ${\bm{m}}^a$ of the thin layer. 

Note that at increasing field the resistivity graph moves rightwards towards higher currents. This is due to the fact that the precession
frequency increases proportionally to the applied field, so that the damping $\Gamma_{a+}=\approx 2\alpha_a\omega_a$ also increases and 
a higher current is needed to compensate it.

\begin{figure}
\begin{center}
\includegraphics[width=8.0cm]{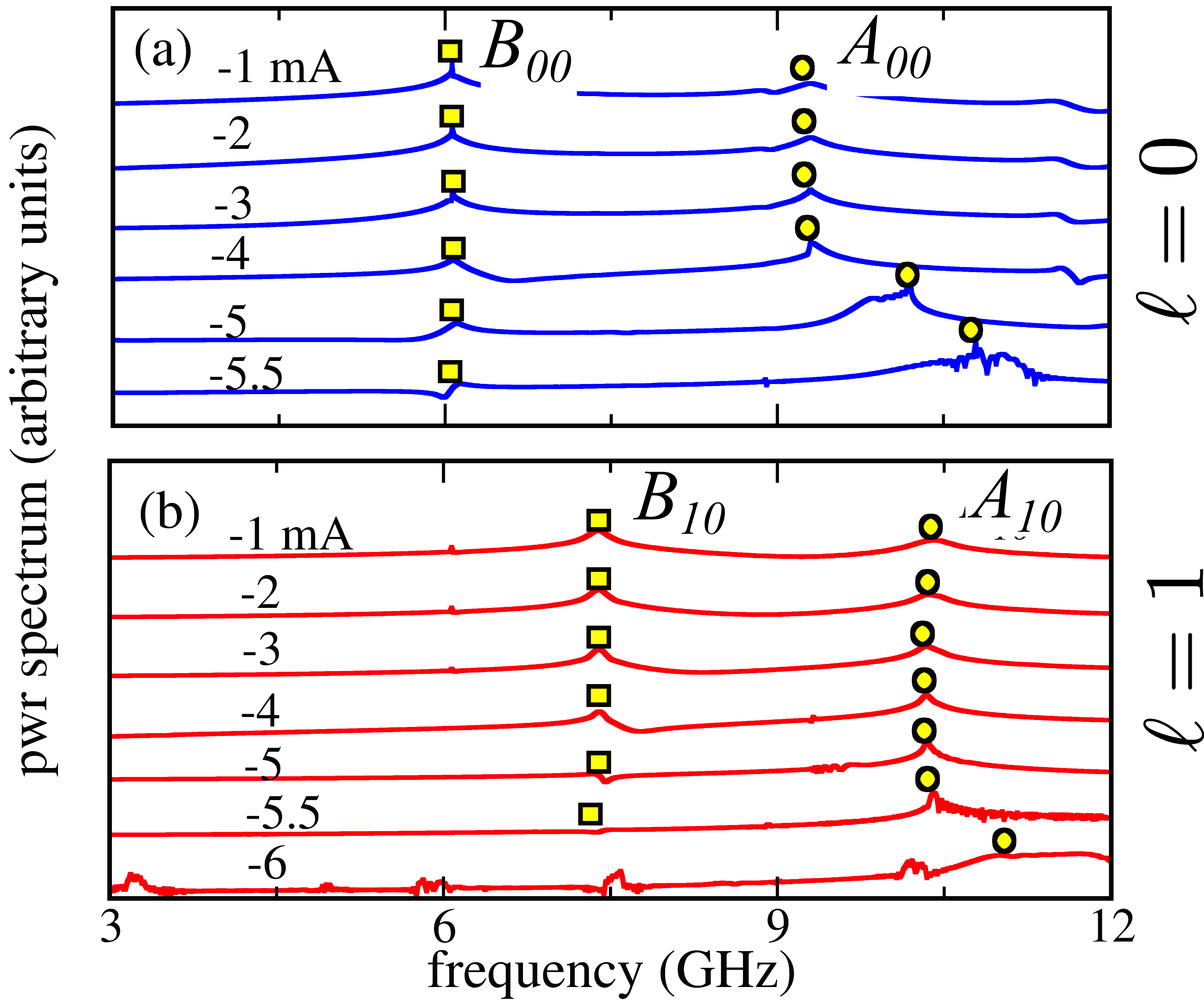}
\caption{SW power spectrum of the $\ell=0$ (a) and $\ell=+1$ (b) modes, computed for different values of the dc current. The yellow squares
(resp. circles) indicate the positions of the $B$ (resp. $A$) modes.} 
\label{fig:result3}
\end{center}
\end{figure}

The power spectrum of the system, with the modes $\ell=0$ and $+1$, is shown respectively in Fig.\ref{fig:result3} (a) and (b) in logarithmic scale. The modes $A_{00}$ and $A_{10}$, corresponding to the dynamics of the thin layer,
increase with the current and dominate the spectrum around $I_{dc}=-5$ mA. Beyond the critical threshold, their frequency increases with the field. On the other hand, the modes $B_{00}$ and $B_{10}$, which dominate the spectrum
at zero currents, decrease until they almost disappear near the critical threshold.

\begin{figure}
\begin{center}
\includegraphics[width=8.0cm]{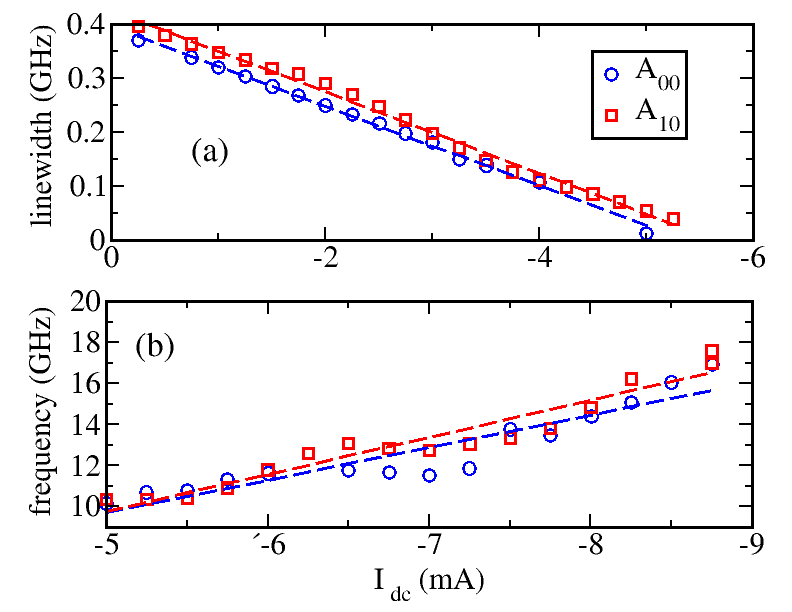}
\caption{Linewidths (a) and frequencies (b) of the modes $A_{00}$ and $A_{10}$ as a function of the dc current near the critical threshold.
The symbols are numerical calculations, while dashed lines are linear fits.} 
\label{fig:result4}
\end{center}
\end{figure}

Fig.\ref{fig:result4} shows the linewidth and frequencies of the modes $A_{00}$ and $A_{10}$ as a function of the dc current. In panel (a), one can see that the linewidths of both modes decrease linearly of an order of magnitude
at increasing current, and they vanish at the critical threshold, where the damping is compensated and the system begins to auto oscillate. 
Note that the two modes have slightly different critical currents, of about $5$ ($\ell=0$) and $5.5$ ($\ell=1$)mA.
The behaviour of the spectrum and the critical currents agree with the experimental result of Refs.[\onlinecite{thesis,naletov11}].
The frequencies of the excited modes remain constant until the critical threshold, and then starts increasing linearly with a rate around 2 GHz per mA. This behaviour also agrees with experimental data.

\begin{figure}
\begin{center}
\includegraphics[width=8.0cm]{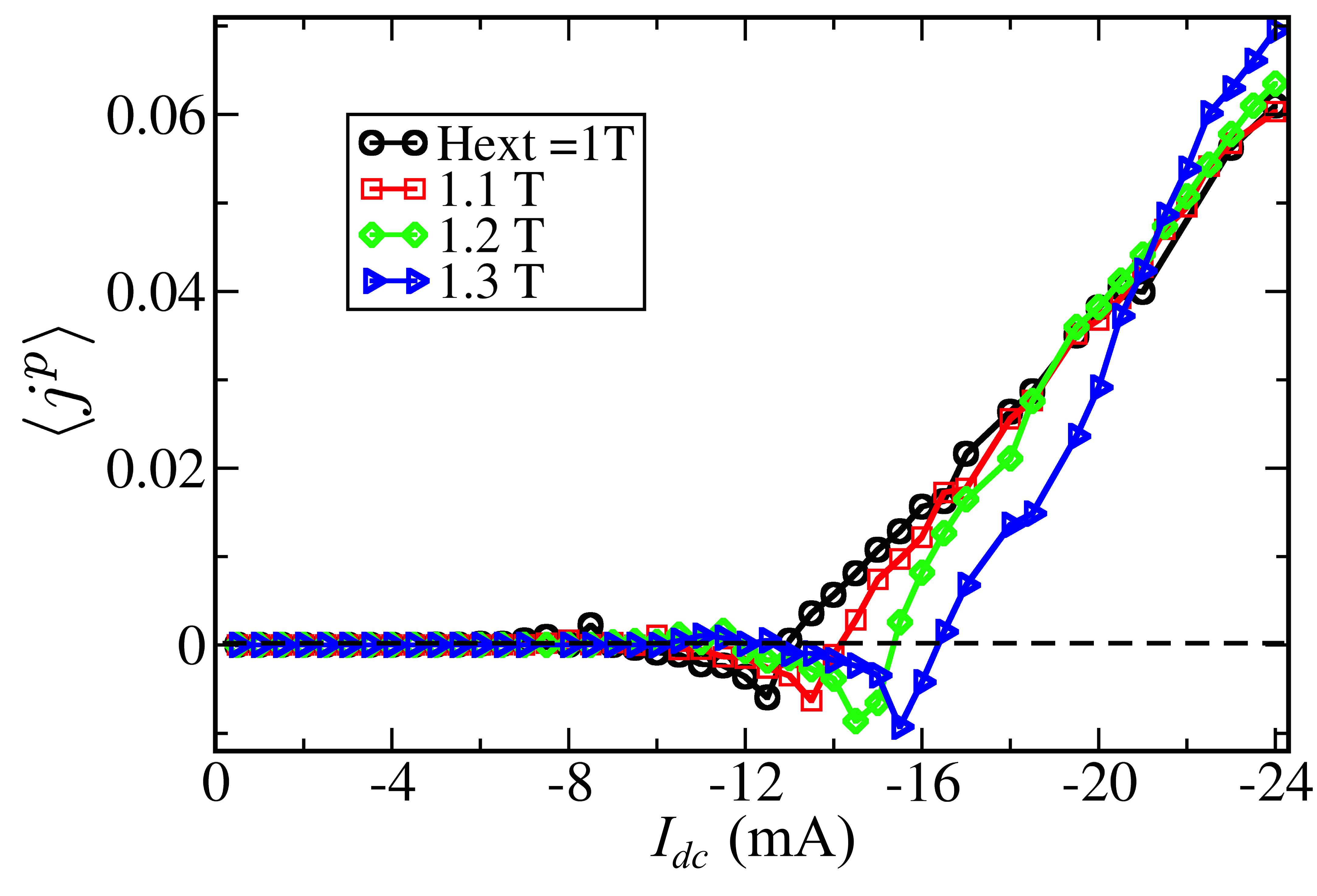}
\caption{Magnetisation current from thick to thin layer as a function of dc current $I_{dc}$, computed for different values of the applied field.}
\label{fig:result5}
\end{center}
\end{figure}
We conclude this section by noting that the physics discussed here is very similar to that of the spin-caloritronics diode \cite{borlenghi14a}. 
In Fig.\ref{fig:result5}, one can see that increasing the electrical dc current leads to an increase of the magnon current between the two disks. This current describes the transfer of
magnetic moment $M_z$ between the two disks, and corresponds to the usual SW current written for a system of only two spins 
\cite{borlenghi14a,borlenghi14b,borlenghi15a,borlenghi15b}.The increase of the SW current with the dc current is due to the fact that spin transfer torque excites only one mode, which eventually dominates the spectrum, inducing a phase locking between the two disks. In this kind of discrete systems, the magnon current is indeed a measure of the phase 
synchronisation of the system.

\section{Conclusions}\label{sec:conclusions}%

By coupling CRMT to micromagnetic simulations with the Nmag code we  allow description, on an equal footing and without free parameters, of both transport and magnetic degrees of freedom. The results of this work have important consequences regarding the characterization  and the optimization of the performance of STNOs.

Using the developed method, we have identified the nature of the modes that auto-oscillate when the current exceeds the critical threshold. In particular, we have predicted different critical currents for the modes $A_{00}$ and $A_{10}$, 
and a nonlinear frequency shift of the order of 2~GHz per mA.
The precise determination of the SW mode symmetry of the auto-oscillating mode is important for the phase synchronization of a STNO to an external source. In fact, it will be successful only if the latter can couple efficiently to the spin transfer driven auto-oscillation modes, 
\textit{i.e.}, if it has the appropriate symmetry.

The flexibility of CRMT and of the procedure to couple it with Nmag allows one to use our numerical
tool to simulate different geometries and materials.
Further development and investigations are possible. A multiscale approach which combines systematically CRMT with a fully quantum approach has been implemented
\cite{borlenghi11}. This should allow to compute current driven dynamics in a large variety of systems, including multi-terminal devices and tunnel junctions. 

We remark that in our simulations we have not taken into account the lateral diffusion of spins, since the system is described using 
one dimensional CRMT columns, where electrons propagate only along the $z$ direction. 
 
This 1D model of transport is effective to describe selection rules into a perpendicularly magnetized nanopillar with magnetic 
field applied along ${\bm e}_z$, but more complicated configurations (such as magnetic vortexes and multi-terminal
spin valves) may require a fully three dimensional description.

Finally, further investigation is necessary to understand the behaviour of the system at high current, where nonlinear effects 
(such as the dependence of the Gilbert damping on current \cite{slavin09}) may play an important role.
The present work can be considered as an intermediate step towards a fully 3D description of transport and magnetization dynamics in realistic systems.

\section{Acknowledgements}%
We wish to thank R. Lassalle-Balier and J. Dubois for fruitful discussions, X. Waintal, O. Klein and G. de Loubens for useful comments and assistance
in the analysis of the experiments.
Financial support from 
Vetenskapsrådet (VR), 
Carl Tryggers Stiftelse (CTS), Stiftelsen Olle Engqvist Byggmästare and Swedish Energy Agency (STEM) is gratefully acknowledged.

The computations were performed on resources provided by the Swedish National Infrastructure for Computing (SNIC) at the National Supercomputer Center (NSC), Link\"oping University, the PDC Centre for High Performance Computing (PDC-HPC), KTH, and the High Performance Computing Center North  (HPC2N), Umeå University.



\begin{thebibliography}{38}
\expandafter\ifx\csname natexlab\endcsname\relax\def\natexlab#1{#1}\fi
\expandafter\ifx\csname bibnamefont\endcsname\relax
  \def\bibnamefont#1{#1}\fi
\expandafter\ifx\csname bibfnamefont\endcsname\relax
  \def\bibfnamefont#1{#1}\fi
\expandafter\ifx\csname citenamefont\endcsname\relax
  \def\citenamefont#1{#1}\fi
\expandafter\ifx\csname url\endcsname\relax
  \def\url#1{\texttt{#1}}\fi
\expandafter\ifx\csname urlprefix\endcsname\relax\def\urlprefix{URL }\fi
\providecommand{\bibinfo}[2]{#2}
\providecommand{\eprint}[2][]{\url{#2}}

\bibitem[{\citenamefont{Berger}(1978)}]{berger78}
\bibinfo{author}{\bibfnamefont{L.}~\bibnamefont{Berger}},
  \bibinfo{journal}{Journal of Applied Physics} \textbf{\bibinfo{volume}{49}},
  \bibinfo{pages}{2156} (\bibinfo{year}{1978}).

\bibitem[{\citenamefont{Berger}(1996)}]{berger96}
\bibinfo{author}{\bibfnamefont{L.}~\bibnamefont{Berger}},
  \bibinfo{journal}{Phys. Rev. B} \textbf{\bibinfo{volume}{54}},
  \bibinfo{pages}{9353} (\bibinfo{year}{1996}).

\bibitem[{\citenamefont{Slonczewski}(1996)}]{slonczewski96}
\bibinfo{author}{\bibfnamefont{J.~C.} \bibnamefont{Slonczewski}},
  \bibinfo{journal}{Journal of Magnetism and Magnetic Materials}
  \textbf{\bibinfo{volume}{159}}, \bibinfo{pages}{L1 } (\bibinfo{year}{1996}),
  ISSN \bibinfo{issn}{0304-8853}.

\bibitem[{\citenamefont{Baibich et~al.}(1988)\citenamefont{Baibich, Broto,
  Fert, Van~Dau, Petroff, Etienne, Creuzet, Friederich, and
  Chazelas}}]{baibich88}
\bibinfo{author}{\bibfnamefont{M.~N.} \bibnamefont{Baibich}},
  \bibinfo{author}{\bibfnamefont{J.~M.} \bibnamefont{Broto}},
  \bibinfo{author}{\bibfnamefont{A.}~\bibnamefont{Fert}},
  \bibinfo{author}{\bibfnamefont{F.~N.} \bibnamefont{Van~Dau}},
  \bibinfo{author}{\bibfnamefont{F.}~\bibnamefont{Petroff}},
  \bibinfo{author}{\bibfnamefont{P.}~\bibnamefont{Etienne}},
  \bibinfo{author}{\bibfnamefont{G.}~\bibnamefont{Creuzet}},
  \bibinfo{author}{\bibfnamefont{A.}~\bibnamefont{Friederich}},
  \bibnamefont{and} \bibinfo{author}{\bibfnamefont{J.}~\bibnamefont{Chazelas}},
  \bibinfo{journal}{Phys. Rev. Lett.} \textbf{\bibinfo{volume}{61}},
  \bibinfo{pages}{2472} (\bibinfo{year}{1988}).

\bibitem[{\citenamefont{Binasch et~al.}(1989)\citenamefont{Binasch, Gr\"unberg,
  Saurenbach, and Zinn}}]{binash89}
\bibinfo{author}{\bibfnamefont{G.}~\bibnamefont{Binasch}},
  \bibinfo{author}{\bibfnamefont{P.}~\bibnamefont{Gr\"unberg}},
  \bibinfo{author}{\bibfnamefont{F.}~\bibnamefont{Saurenbach}},
  \bibnamefont{and} \bibinfo{author}{\bibfnamefont{W.}~\bibnamefont{Zinn}},
  \bibinfo{journal}{Phys. Rev. B} \textbf{\bibinfo{volume}{39}},
  \bibinfo{pages}{4828} (\bibinfo{year}{1989}).

\bibitem[{\citenamefont{Diao et~al.}(2007)\citenamefont{Diao, Li, Wang, Ding,
  Panchula, Chen, Wang, and Huai}}]{diao07}
\bibinfo{author}{\bibfnamefont{Z.}~\bibnamefont{Diao}},
  \bibinfo{author}{\bibfnamefont{Z.}~\bibnamefont{Li}},
  \bibinfo{author}{\bibfnamefont{S.}~\bibnamefont{Wang}},
  \bibinfo{author}{\bibfnamefont{Y.}~\bibnamefont{Ding}},
  \bibinfo{author}{\bibfnamefont{A.}~\bibnamefont{Panchula}},
  \bibinfo{author}{\bibfnamefont{E.}~\bibnamefont{Chen}},
  \bibinfo{author}{\bibfnamefont{L.-C.} \bibnamefont{Wang}}, \bibnamefont{and}
  \bibinfo{author}{\bibfnamefont{Y.}~\bibnamefont{Huai}},
  \bibinfo{journal}{Journal of Physics: Condensed Matter}
  \textbf{\bibinfo{volume}{19}}, \bibinfo{pages}{165209}
  (\bibinfo{year}{2007}).

\bibitem[{\citenamefont{Katine et~al.}(2000)\citenamefont{Katine, Albert,
  Buhrman, Myers, and Ralph}}]{katine00}
\bibinfo{author}{\bibfnamefont{J.~A.} \bibnamefont{Katine}},
  \bibinfo{author}{\bibfnamefont{F.~J.} \bibnamefont{Albert}},
  \bibinfo{author}{\bibfnamefont{R.~A.} \bibnamefont{Buhrman}},
  \bibinfo{author}{\bibfnamefont{E.~B.} \bibnamefont{Myers}}, \bibnamefont{and}
  \bibinfo{author}{\bibfnamefont{D.~C.} \bibnamefont{Ralph}},
  \bibinfo{journal}{Phys. Rev. Lett.} \textbf{\bibinfo{volume}{84}},
  \bibinfo{pages}{3149} (\bibinfo{year}{2000}).

\bibitem[{\citenamefont{C. et~al.}(2014)\citenamefont{C., M., F., andHrkac
  G.~amd Praetorius~D., and D.}}]{abert14}
\bibinfo{author}{\bibfnamefont{A.}~\bibnamefont{C.}},
  \bibinfo{author}{\bibfnamefont{R.}~\bibnamefont{M.}},
  \bibinfo{author}{\bibfnamefont{B.}~\bibnamefont{F.}},
  \bibinfo{author}{\bibfnamefont{V.~C.} \bibnamefont{andHrkac G.~amd
  Praetorius~D.}}, \bibnamefont{and}
  \bibinfo{author}{\bibfnamefont{S.}~\bibnamefont{D.}}, Ph.D. thesis
  (\bibinfo{year}{2014}).

\bibitem[{\citenamefont{K.~J.~Lee}(2004)}]{lee04}
\bibinfo{author}{\bibfnamefont{O.~R. J. P. N. B.~D.} \bibnamefont{K.~J.~Lee},
  \bibfnamefont{A.~Deac}}, \bibinfo{journal}{Nat. Mater.}
  \textbf{\bibinfo{volume}{3}}, \bibinfo{pages}{877} (\bibinfo{year}{2004}).

\bibitem[{\citenamefont{Xiao et~al.}(2005)\citenamefont{Xiao, Zangwill, and
  Stiles}}]{xiao05}
\bibinfo{author}{\bibfnamefont{J.}~\bibnamefont{Xiao}},
  \bibinfo{author}{\bibfnamefont{A.}~\bibnamefont{Zangwill}}, \bibnamefont{and}
  \bibinfo{author}{\bibfnamefont{M.~D.} \bibnamefont{Stiles}},
  \bibinfo{journal}{Phys. Rev. B} \textbf{\bibinfo{volume}{72}},
  \bibinfo{pages}{014446} (\bibinfo{year}{2005}).

\bibitem[{\citenamefont{Berkov and Miltat}(2008)}]{berkov08}
\bibinfo{author}{\bibfnamefont{D.}~\bibnamefont{Berkov}} \bibnamefont{and}
  \bibinfo{author}{\bibfnamefont{J.}~\bibnamefont{Miltat}},
  \bibinfo{journal}{Journal of Magnetism and Magnetic Materials}
  \textbf{\bibinfo{volume}{320}}, \bibinfo{pages}{1238 }
  (\bibinfo{year}{2008}), ISSN \bibinfo{issn}{0304-8853}.

\bibitem[{\citenamefont{Fischbacher et~al.}(2007)\citenamefont{Fischbacher,
  Franchin, Bordignon, and Fangohr}}]{fischbacher07}
\bibinfo{author}{\bibfnamefont{T.}~\bibnamefont{Fischbacher}},
  \bibinfo{author}{\bibfnamefont{M.}~\bibnamefont{Franchin}},
  \bibinfo{author}{\bibfnamefont{G.}~\bibnamefont{Bordignon}},
  \bibnamefont{and} \bibinfo{author}{\bibfnamefont{H.}~\bibnamefont{Fangohr}},
  \bibinfo{journal}{IEEE Trans. Magn.} \textbf{\bibinfo{volume}{43}},
  \bibinfo{pages}{2896} (\bibinfo{year}{2007}).

\bibitem[{\citenamefont{Rychkov et~al.}(2009)\citenamefont{Rychkov, Borlenghi,
  Jaffres, Fert, and Waintal}}]{rychkov09}
\bibinfo{author}{\bibfnamefont{V.~S.} \bibnamefont{Rychkov}},
  \bibinfo{author}{\bibfnamefont{S.}~\bibnamefont{Borlenghi}},
  \bibinfo{author}{\bibfnamefont{H.}~\bibnamefont{Jaffres}},
  \bibinfo{author}{\bibfnamefont{A.}~\bibnamefont{Fert}}, \bibnamefont{and}
  \bibinfo{author}{\bibfnamefont{X.}~\bibnamefont{Waintal}},
  \bibinfo{journal}{Phys. Rev. Lett.} \textbf{\bibinfo{volume}{103}},
  \bibinfo{pages}{066602} (\bibinfo{year}{2009}).

\bibitem[{\citenamefont{Borlenghi et~al.}(2011)\citenamefont{Borlenghi,
  Rychkov, Petitjean, and Waintal}}]{borlenghi11}
\bibinfo{author}{\bibfnamefont{S.}~\bibnamefont{Borlenghi}},
  \bibinfo{author}{\bibfnamefont{V.}~\bibnamefont{Rychkov}},
  \bibinfo{author}{\bibfnamefont{C.}~\bibnamefont{Petitjean}},
  \bibnamefont{and} \bibinfo{author}{\bibfnamefont{X.}~\bibnamefont{Waintal}},
  \bibinfo{journal}{Phys. Rev. B} \textbf{\bibinfo{volume}{84}},
  \bibinfo{pages}{035412} (\bibinfo{year}{2011}).

\bibitem[{\citenamefont{Valet and Fert}(1993)}]{valet93}
\bibinfo{author}{\bibfnamefont{T.}~\bibnamefont{Valet}} \bibnamefont{and}
  \bibinfo{author}{\bibfnamefont{A.}~\bibnamefont{Fert}},
  \bibinfo{journal}{Phys. Rev. B} \textbf{\bibinfo{volume}{48}},
  \bibinfo{pages}{7099} (\bibinfo{year}{1993}).

\bibitem[{\citenamefont{Naletov et~al.}(2011)}]{naletov11}
\bibinfo{author}{\bibfnamefont{V.~V.} \bibnamefont{Naletov}}
  \bibnamefont{et~al.}, \bibinfo{journal}{Phys. Rev. B}
  \textbf{\bibinfo{volume}{84}}, \bibinfo{pages}{224423}
  (\bibinfo{year}{2011}).

\bibitem[{\citenamefont{Borlenghi
  et~al.}(2014{\natexlab{a}})\citenamefont{Borlenghi, Wang, Fangohr, Bergqvist,
  and Delin}}]{borlenghi14a}
\bibinfo{author}{\bibfnamefont{S.}~\bibnamefont{Borlenghi}},
  \bibinfo{author}{\bibfnamefont{W.}~\bibnamefont{Wang}},
  \bibinfo{author}{\bibfnamefont{H.}~\bibnamefont{Fangohr}},
  \bibinfo{author}{\bibfnamefont{L.}~\bibnamefont{Bergqvist}},
  \bibnamefont{and} \bibinfo{author}{\bibfnamefont{A.}~\bibnamefont{Delin}},
  \bibinfo{journal}{Phys. Rev. Lett.} \textbf{\bibinfo{volume}{112}},
  \bibinfo{pages}{047203} (\bibinfo{year}{2014}{\natexlab{a}}).

\bibitem[{\citenamefont{Borlenghi
  et~al.}(2014{\natexlab{b}})\citenamefont{Borlenghi, Lepri, Bergqvist, and
  Delin}}]{borlenghi14b}
\bibinfo{author}{\bibfnamefont{S.}~\bibnamefont{Borlenghi}},
  \bibinfo{author}{\bibfnamefont{S.}~\bibnamefont{Lepri}},
  \bibinfo{author}{\bibfnamefont{L.}~\bibnamefont{Bergqvist}},
  \bibnamefont{and} \bibinfo{author}{\bibfnamefont{A.}~\bibnamefont{Delin}},
  \bibinfo{journal}{Phys. Rev. B} \textbf{\bibinfo{volume}{89}},
  \bibinfo{pages}{054428} (\bibinfo{year}{2014}{\natexlab{b}}).

\bibitem[{\citenamefont{Borlenghi
  et~al.}(2015{\natexlab{a}})\citenamefont{Borlenghi, Iubini, Lepri, Bergqvist,
  Delin, and Fransson}}]{borlenghi15a}
\bibinfo{author}{\bibfnamefont{S.}~\bibnamefont{Borlenghi}},
  \bibinfo{author}{\bibfnamefont{S.}~\bibnamefont{Iubini}},
  \bibinfo{author}{\bibfnamefont{S.}~\bibnamefont{Lepri}},
  \bibinfo{author}{\bibfnamefont{L.}~\bibnamefont{Bergqvist}},
  \bibinfo{author}{\bibfnamefont{A.}~\bibnamefont{Delin}}, \bibnamefont{and}
  \bibinfo{author}{\bibfnamefont{J.}~\bibnamefont{Fransson}},
  \bibinfo{journal}{Phys. Rev. E} \textbf{\bibinfo{volume}{91}},
  \bibinfo{pages}{040102} (\bibinfo{year}{2015}{\natexlab{a}}).

\bibitem[{\citenamefont{Borlenghi
  et~al.}(2015{\natexlab{b}})\citenamefont{Borlenghi, Iubini, Lepri, Chico,
  Bergqvist, Delin, and Fransson}}]{borlenghi15b}
\bibinfo{author}{\bibfnamefont{S.}~\bibnamefont{Borlenghi}},
  \bibinfo{author}{\bibfnamefont{S.}~\bibnamefont{Iubini}},
  \bibinfo{author}{\bibfnamefont{S.}~\bibnamefont{Lepri}},
  \bibinfo{author}{\bibfnamefont{J.}~\bibnamefont{Chico}},
  \bibinfo{author}{\bibfnamefont{L.}~\bibnamefont{Bergqvist}},
  \bibinfo{author}{\bibfnamefont{A.}~\bibnamefont{Delin}}, \bibnamefont{and}
  \bibinfo{author}{\bibfnamefont{J.}~\bibnamefont{Fransson}},
  \bibinfo{journal}{Phys. Rev. E} \textbf{\bibinfo{volume}{92}},
  \bibinfo{pages}{012116} (\bibinfo{year}{2015}{\natexlab{b}}).

\bibitem[{\citenamefont{Slavin and Tiberkevich}(2009)}]{slavin09}
\bibinfo{author}{\bibfnamefont{A.}~\bibnamefont{Slavin}} \bibnamefont{and}
  \bibinfo{author}{\bibfnamefont{V.}~\bibnamefont{Tiberkevich}},
  \bibinfo{journal}{Magnetics, IEEE Transactions on}
  \textbf{\bibinfo{volume}{45}}, \bibinfo{pages}{1875 } (\bibinfo{year}{2009}),
  ISSN \bibinfo{issn}{0018-9464}.

\bibitem[{\citenamefont{Waintal et~al.}(2000)\citenamefont{Waintal, Myers,
  Brouwer, and Ralph}}]{waintal00}
\bibinfo{author}{\bibfnamefont{X.}~\bibnamefont{Waintal}},
  \bibinfo{author}{\bibfnamefont{E.~B.} \bibnamefont{Myers}},
  \bibinfo{author}{\bibfnamefont{P.~W.} \bibnamefont{Brouwer}},
  \bibnamefont{and} \bibinfo{author}{\bibfnamefont{D.~C.} \bibnamefont{Ralph}},
  \bibinfo{journal}{Phys. Rev. B} \textbf{\bibinfo{volume}{62}},
  \bibinfo{pages}{12317} (\bibinfo{year}{2000}).

\bibitem[{\citenamefont{Gilbert}(2004)}]{gilbert55}
\bibinfo{author}{\bibfnamefont{T.}~\bibnamefont{Gilbert}},
  \bibinfo{journal}{Magnetics, IEEE Transactions on}
  \textbf{\bibinfo{volume}{40}}, \bibinfo{pages}{3443 } (\bibinfo{year}{2004}).

\bibitem[{\citenamefont{Landau and Lifshitz}(1965)}]{landau65}
\bibinfo{author}{\bibfnamefont{L.~D.} \bibnamefont{Landau}} \bibnamefont{and}
  \bibinfo{author}{\bibfnamefont{E.~M.} \bibnamefont{Lifshitz}}, in
  \emph{\bibinfo{booktitle}{Collected paper}} (\bibinfo{publisher}{Ed.
  Pergamon}, \bibinfo{year}{1965}).

\bibitem[{\citenamefont{Gurevich and Melkov}(1996)}]{gurevich96}
\bibinfo{author}{\bibfnamefont{A.~G.} \bibnamefont{Gurevich}} \bibnamefont{and}
  \bibinfo{author}{\bibfnamefont{G.~A.} \bibnamefont{Melkov}},
  \emph{\bibinfo{title}{Magnetization {O}scillation and {W}aves}}
  (\bibinfo{publisher}{CRC Press}, \bibinfo{year}{1996}).

\bibitem[{\citenamefont{Sankey et~al.}(2006)\citenamefont{Sankey, Braganca,
  Garcia, Krivorotov, Buhrman, and Ralph}}]{sankey06}
\bibinfo{author}{\bibfnamefont{J.~C.} \bibnamefont{Sankey}},
  \bibinfo{author}{\bibfnamefont{P.~M.} \bibnamefont{Braganca}},
  \bibinfo{author}{\bibfnamefont{A.~G.~F.} \bibnamefont{Garcia}},
  \bibinfo{author}{\bibfnamefont{I.~N.} \bibnamefont{Krivorotov}},
  \bibinfo{author}{\bibfnamefont{R.~A.} \bibnamefont{Buhrman}},
  \bibnamefont{and} \bibinfo{author}{\bibfnamefont{D.~C.} \bibnamefont{Ralph}},
  \bibinfo{journal}{Phys. Rev. Lett.} \textbf{\bibinfo{volume}{96}},
  \bibinfo{eid}{227601} (\bibinfo{year}{2006}).

\bibitem[{\citenamefont{Chen et~al.}(2008)\citenamefont{Chen, Beaujour,
  de~Loubens, Kent, and Sun}}]{chen08}
\bibinfo{author}{\bibfnamefont{W.}~\bibnamefont{Chen}},
  \bibinfo{author}{\bibfnamefont{J.-M.~L.} \bibnamefont{Beaujour}},
  \bibinfo{author}{\bibfnamefont{G.}~\bibnamefont{de~Loubens}},
  \bibinfo{author}{\bibfnamefont{A.~D.} \bibnamefont{Kent}}, \bibnamefont{and}
  \bibinfo{author}{\bibfnamefont{J.~Z.} \bibnamefont{Sun}},
  \bibinfo{journal}{Appl. Phys. Lett.} \textbf{\bibinfo{volume}{92}},
  \bibinfo{eid}{012507} (\bibinfo{year}{2008}).

\bibitem[{\citenamefont{Iubini et~al.}(2013)\citenamefont{Iubini, Lepri, Livi,
  and Politi}}]{iubini13}
\bibinfo{author}{\bibfnamefont{S.}~\bibnamefont{Iubini}},
  \bibinfo{author}{\bibfnamefont{S.}~\bibnamefont{Lepri}},
  \bibinfo{author}{\bibfnamefont{R.}~\bibnamefont{Livi}}, \bibnamefont{and}
  \bibinfo{author}{\bibfnamefont{A.}~\bibnamefont{Politi}},
  \bibinfo{journal}{Journal of Statistical Mechanics: Theory and Experiment}
  \textbf{\bibinfo{volume}{2013}}, \bibinfo{pages}{P08017}
  (\bibinfo{year}{2013}).

\bibitem[{\citenamefont{Damon and Eshbach}(1961)}]{damon61}
\bibinfo{author}{\bibfnamefont{R.~W.} \bibnamefont{Damon}} \bibnamefont{and}
  \bibinfo{author}{\bibfnamefont{J.~R.} \bibnamefont{Eshbach}},
  \bibinfo{journal}{J. Phys. Chem. Solids} \textbf{\bibinfo{volume}{19}},
  \bibinfo{pages}{308} (\bibinfo{year}{1961}).

\bibitem[{\citenamefont{Belmeguenai et~al.}(2007)\citenamefont{Belmeguenai,
  Martin, Woltersdorf, Maier, and Bayreuther}}]{belmeguenai07}
\bibinfo{author}{\bibfnamefont{M.}~\bibnamefont{Belmeguenai}},
  \bibinfo{author}{\bibfnamefont{T.}~\bibnamefont{Martin}},
  \bibinfo{author}{\bibfnamefont{G.}~\bibnamefont{Woltersdorf}},
  \bibinfo{author}{\bibfnamefont{M.}~\bibnamefont{Maier}}, \bibnamefont{and}
  \bibinfo{author}{\bibfnamefont{G.}~\bibnamefont{Bayreuther}},
  \bibinfo{journal}{Phys. Rev. B} \textbf{\bibinfo{volume}{76}},
  \bibinfo{pages}{104414} (\bibinfo{year}{2007}).

\bibitem[{\citenamefont{Gubbiotti et~al.}(2004)\citenamefont{Gubbiotti,
  Kostylev, Sergeeva, Conti, Carlotti, Ono, Slavin, and
  Stashkevich}}]{gubbiotti04}
\bibinfo{author}{\bibfnamefont{G.}~\bibnamefont{Gubbiotti}},
  \bibinfo{author}{\bibfnamefont{M.}~\bibnamefont{Kostylev}},
  \bibinfo{author}{\bibfnamefont{N.}~\bibnamefont{Sergeeva}},
  \bibinfo{author}{\bibfnamefont{M.}~\bibnamefont{Conti}},
  \bibinfo{author}{\bibfnamefont{G.}~\bibnamefont{Carlotti}},
  \bibinfo{author}{\bibfnamefont{T.}~\bibnamefont{Ono}},
  \bibinfo{author}{\bibfnamefont{A.~N.} \bibnamefont{Slavin}},
  \bibnamefont{and}
  \bibinfo{author}{\bibfnamefont{A.}~\bibnamefont{Stashkevich}},
  \bibinfo{journal}{Phys. Rev. B} \textbf{\bibinfo{volume}{70}},
  \bibinfo{pages}{224422} (\bibinfo{year}{2004}).

\bibitem[{\citenamefont{Hindmarsh et~al.}(2005)\citenamefont{Hindmarsh, Brown,
  Grant, Lee, Serban, Shumaker, and Woodward}}]{hindmarsh05}
\bibinfo{author}{\bibfnamefont{A.~C.} \bibnamefont{Hindmarsh}},
  \bibinfo{author}{\bibfnamefont{P.~N.} \bibnamefont{Brown}},
  \bibinfo{author}{\bibfnamefont{K.~E.} \bibnamefont{Grant}},
  \bibinfo{author}{\bibfnamefont{S.~L.} \bibnamefont{Lee}},
  \bibinfo{author}{\bibfnamefont{R.}~\bibnamefont{Serban}},
  \bibinfo{author}{\bibfnamefont{D.~E.} \bibnamefont{Shumaker}},
  \bibnamefont{and} \bibinfo{author}{\bibfnamefont{C.~S.}
  \bibnamefont{Woodward}}, \bibinfo{journal}{ACM Trans. Math. Softw.}
  \textbf{\bibinfo{volume}{31}}, \bibinfo{pages}{363} (\bibinfo{year}{2005}),
  ISSN \bibinfo{issn}{0098-3500}.

\bibitem[{\citenamefont{B\"uttiker et~al.}(1985)\citenamefont{B\"uttiker, Imry,
  Landauer, and Pinhas}}]{buttiker85}
\bibinfo{author}{\bibfnamefont{M.}~\bibnamefont{B\"uttiker}},
  \bibinfo{author}{\bibfnamefont{Y.}~\bibnamefont{Imry}},
  \bibinfo{author}{\bibfnamefont{R.}~\bibnamefont{Landauer}}, \bibnamefont{and}
  \bibinfo{author}{\bibfnamefont{S.}~\bibnamefont{Pinhas}},
  \bibinfo{journal}{Phys. Rev. B} \textbf{\bibinfo{volume}{31}},
  \bibinfo{pages}{6207} (\bibinfo{year}{1985}).

\bibitem[{\citenamefont{Akkermans and Montambaux}(2007)}]{montambaux07}
\bibinfo{author}{\bibfnamefont{E.}~\bibnamefont{Akkermans}} \bibnamefont{and}
  \bibinfo{author}{\bibfnamefont{G.}~\bibnamefont{Montambaux}},
  \emph{\bibinfo{title}{Mesoscopic physics of electrons and photons}}
  (\bibinfo{publisher}{Cambridge University Press}, \bibinfo{year}{2007}).

\bibitem[{\citenamefont{Beenakker}(1997)}]{beenakker97}
\bibinfo{author}{\bibfnamefont{C.~W.~J.} \bibnamefont{Beenakker}},
  \bibinfo{journal}{Rev. Mod. Phys.} \textbf{\bibinfo{volume}{69}},
  \bibinfo{pages}{731} (\bibinfo{year}{1997}).

\bibitem[{\citenamefont{Mehta}(2004)}]{mehta04}
\bibinfo{author}{\bibfnamefont{M.~L.} \bibnamefont{Mehta}},
  \emph{\bibinfo{title}{Random Matrices}} (\bibinfo{publisher}{Academic Press},
  \bibinfo{year}{2004}).

\bibitem[{\citenamefont{Petitjean et~al.}(2012)\citenamefont{Petitjean, Luc,
  and Waintal}}]{petitjean12}
\bibinfo{author}{\bibfnamefont{C.}~\bibnamefont{Petitjean}},
  \bibinfo{author}{\bibfnamefont{D.}~\bibnamefont{Luc}}, \bibnamefont{and}
  \bibinfo{author}{\bibfnamefont{X.}~\bibnamefont{Waintal}},
  \bibinfo{journal}{Phys. Rev. Lett.} \textbf{\bibinfo{volume}{109}},
  \bibinfo{pages}{117204} (\bibinfo{year}{2012}).

\bibitem[{\citenamefont{Borlenghi}(2011)}]{thesis}
\bibinfo{author}{\bibfnamefont{S.}~\bibnamefont{Borlenghi}},
  \bibinfo{type}{Theses}, \bibinfo{school}{{Universit{\'e} Pierre et Marie
  Curie - Paris VI}} (\bibinfo{year}{2011}).

\end{thebibliography}

\end{document}